\documentclass[11pt]{article}
\usepackage[symbol]{footmisc}
\def\correspondingauthor{\footnote{Corresponding author.  }}
\usepackage[left=2cm,top=2.5cm,right=2cm,bottom=2.5cm]{geometry}
\usepackage{amsmath, amssymb}

\usepackage{graphicx}
\usepackage{graphics}
\usepackage{epstopdf}
\usepackage{subfigure}
\usepackage{amsfonts}
\usepackage{sectsty}
\usepackage{sectsty}
\usepackage{hyperref}
\usepackage{cite}

\begin{document}

	\begin{center}
	\large{\bf{Traversable Wormholes with Exponential Shape Function in Modified Gravity and General Relativity: A Comparative Study  }} \\
	\vspace{5mm}
	\normalsize{Gauranga C. Samanta$^1$, Nisha Godani$^{2}$ and Kazuharu Bamba$^{3} {}$\correspondingauthor{} }\\
	\normalsize{$^1$Department of Mathematics, BITS Pilani K K Birla Goa Campus, Goa, India}\\
	$^2$Department of Mathematics, Institute of Applied Sciences and Humanities\\ GLA University, Mathura, Uttar Pradesh, India\\
$^{3}$Division of Human Support System, Faculty of Symbiotic Systems Science,\\
Fukushima University, Fukushima 960-1296, Japan \\
	\normalsize {gauranga81@gmail.com\\nishagodani.dei@gmail.com\\bamba@sss.fukushima-u.ac.jp}
\end{center}
\begin{abstract}
We have proposed a novel shape function on which the metric that models traversable wormholes is dependent. Using this shape function, the energy conditions, equation of state and anisotropy parameter are analyzed in $f(R)$ gravity, $f(R,T)$ gravity and general relativity. Furthermore, the consequences obtained with respect to these theories are compared. In addition, the existence of wormhole geometries is investigated.
\end{abstract}

\section{Literature Survey}
Wormhole solutions of Einstein's field equations in general relativity do not satisfy the classical energy conditions. These connect two universes or two remote parts of the same universe. These were first studied by Flamm \cite{flamm} in a simplest form. Afterwards, Einstein and Rosen \cite{eins-ros}  first introduced a mathematical model of wormhole representing the connection of two asymptotically flat spaces through a bridge  which is known as Einstein-Rosen bridge. It was observed that an exotic matter might present to produce antigravity for stability of the Einstein-Rosen bridge. Otherwise due to gravity, the throat might collapse into a singularity and hence the passage through the wormhole would stop.
According to standard no-go theorems, there could be no stationary traversable wormholes  minimally coupled with physical non-exotic sources in four-dimensional general relativity \cite{FSNLobor, Luke}. Therefore, the existence of wormholes requires either the violation of energy conditions or that the source itself should be non-stationary.
Hence, Beato et al. \cite{Beato} and Canfora et al. \cite{Canfora} constructed an exact and static traversable Lorentzian wormhole in general theory of relativity, minimally coupled to the nonlinear sigma model with a negative cosmological constant. They  found that there was no ``exotic matter" (as a negative cosmological constant could hardly be considered exotic)  and  the wormhole was traversable, Lorentzian and built purely with ingredients arising in standard particle physics. Further, \cite{Beato, Canfora} suggested that exotic matter was not compulsory to construct a traversable wormhole in general theory of relativity.
Morris and Thorne \cite{morris1} made a significant contribution in the study of traversable wormholes, and began an active area of research.  Some examples of traversable wormholes are discussed in
\cite{morris2, Visser1989, Moffat1991, Narahara, visser, Armen}. In literature, various studies have been done to investigate the stability of wormholes and the matter passing through them \cite{Bronnikov1, Shinkai, Gonz, Doro}. 
Lemos  et al. \cite{lemos68} reviewed traversable wormholes, analyzed them due to the effect of cosmological constant and explored various properties.
B\"{o}hmer et al. \cite{Bohmer} obtained various wormhole solutions using a linear relationship between energy density and pressure and explored phantom wormhole geometries. Cataldo and Meza \cite{cataldo} explored wormhole structures filled with matter components of two types.
Cataldo et al. \cite{cataldo1} studied static and spherically symmetric wormholes sustained by matter
sources with isotropic pressures and showed their non-existence in the presence of zero-tidal-force.
Jahromi and Moradpour \cite{moradpour} carried out  a study of static traversable wormholes and found a possibility of a  Lyra displacement vector field so that energy conditions could be satisfied in Lyra manifold.
Barros and Lobo \cite{barros} studied  wormhole structures using three form fields and obtained various solutions. They found the validation of weak and null energy conditions in the presence of three-form fields.

Now a day's, the $f(R)$ and $f(R,T)$ theories have received much attention to describe the several aspects of the universe. In the early 1980s, Starobinsky \cite{Star} discussed $f(R)$ model by taking $f(R)=R+\alpha R^2$, where $\alpha>0$, representing inflationary scenario. Caroll  et al. \cite{caroll} made some corrections in gravitational action by adding the term $R^{-n}$, where $n>0$, and explained cosmic acceleration. Tsujikawa \cite{Tsujikawa} discussed $f(R)$ models  that could satisfy local gravity constraints and the conditions of cosmological viability.
Subsequently, many viable cosmological models in modified $f(R)$ gravity have been discussed in \cite{Capozziello:2002rd, Nojiri, Faraoni, Faul, Cap, Amen}. Moreover, Harko et al. \cite{harko} generalized the gravitational action $f(R)$ by including the stress energy tensor term and developed a new theory called $f(R,T)$ theory of gravity. The choice of stress energy tensor in the action could be induced by exotic imperfect fluid or quantum effects. Since the matter and gravity are coupled, the gravity models are dependent on the source term. Consequently, the path of test particles is deviated from the geodesic path which may produce significant results to explore the universe.
Houndjo \cite{Houndjo} obtained the unification of both deceleration and acceleration phases  without neglecting  matter in the context of $f(R,T)$ gravity.
Baffou et al. \cite{Baffou} studied the cosmological evolution of deceleration and equation of state parameters using $f(R,T)$ gravity. Various cosmological models have been studied in $f(R,T)$ theory of gravity \cite{samanta1, samanta2, moreas, yousaf, Shabini, samanta4, Nisha1}. For reviews  on dark energy problem and modified gravity theories, see e.g.~\cite{Copeland:2006wr, Nojiri:2010wj, DeFelice:2010aj, Capozziello:2010zz, Bamba:2015uma}.


Modified theories have also been used for the exploration of wormhole geometry.
Hochberg et al. \cite{Hochberg} solved semi classical field equations representing wormholes. Nojiri et al. \cite{Nojiri1} used effective equation
method and determined the possibility of the induction of wormholes in early time.
Furey and Bendictis \cite{Furey} considered gravitational action with non-linear powers of Ricci scalar and explored the existence of static wormhole.
Dotti et al. \cite{Dotti} obtained wormhole solutions in higher dimensional gravity.
Bouhmadi-L\'{o}pez et al. \cite{Bouh} assumed a specific form of equation of state, used cut-and-paste approach to match an interior spherically symmetric wormhole solution with an exterior Schwarzschild
solution and analyzed the stability of thin-shell.
Duplessis and Easson \cite{Duplessis} obtained exact traversable wormhole and black hole solutions in scale-free $R^2$ gravity which did not require violation of null energy condition.
Najafi et al. \cite{Najafi} considered an extra space-like dimension, examined its effect on scale factor, shape function and energy density, and explored traversable wormhole in the framework of FLRW model.
Rahaman et al. \cite{Rahaman} constructed traversable wormholes in Finsler geometry. They obtained exact solutions for different choices of shape function, redshift function and equation of state and discussed the characteristics of the wormhole models. 

Lobo and Oliveira \cite{Lobo1} studied traversable wormholes in $f(R)$ gravity. They determined the factors that were  responsible for the dissatisfaction of the null energy condition and supported the wormhole structures. They obtained various solutions by taking various equations of state and considering some particular shape functions with constant redshift function.
 Saiedi and Esfahani \cite{Saiedi}, using constant shape and redshift functions,  obtained wormhole solutions in the background of $f(R)$ gravity and investigated null and weak energy conditions.
 Eiroa and Aguirre \cite{Eiroa1} constructed thin-shell wormholes with charge in $f(R)$ gravity and analyzed their stability under perturbations.  They found that the equation of state of the matter at the throat could  be determined by the junction conditions.
 Furthermore, Eiroa and Aguirre \cite{Eiroa2} studied a family of spherically symmetric Lorentzian wormholes in quadratic $f(R)$ gravity, with a thin shell of matter corresponding to the throat.
   Bahamonde et al. \cite{baha} studied wormholes in $f(R)$ gravity. They constructed a dynamical wormhole and found it asymptotically approaching towards the FLRW universe. Zubair et al. \cite{Zubair:2016cde} investigated static spherically symmetric wormholes in $f(R, T)$ gravity with three different form
   of matter contents. Furthermore, Zubair et al. \cite{Zubair:2017hsq} studied stable wormholes on a noncommutative-geometric background in modified gravity.
   Eiroa and Aguirre \cite{Eiroa3} explored spherical thin shell wormholes in $f(R)$ theory of gravity and obtained the existence of stable static configurations for a suitable set of model parameters. Kuhfittig \cite{peter} explored wormholes in $f(R)$ gravity. He considered various shape functions, derived corresponding $f(R)$ functions and found wormhole solutions. He also considered a special form of function $f(R)$ and obtained wormhole solutions. Godani and Samanta \cite{ns, Godani19} and Samanta and Godani \cite{Samanta19, Samantaepjc}  investigated energy conditions for traversable wormhole in $f(R)$ gravity. Zubair et al. \cite{Zubair:2019yvo, Zubair2019ijmpd} studied static traversable wormhole solution in modified gravity from different aspects.
Moraes et al. \cite{Moraes5} obtained analytical general solutions for  static  wormholes in $f(R,T)$ gravity. Zubair et al. \cite{Zubair} investigated energy conditions and wormhole solutions by taking three types of fluids in $f(R,T)$ gravity.
Moraes and Sahoo \cite{sahoo} studied  static wormholes in $f(R,T)$ theory of gravity and presented some models of wormholes using different assumptions for
the matter content. Recently,  Godani and Samanta \cite{NG_chinese} defined a nonlinear $f(R,T)$ function and explored the spherical regions for static traversable wormholes  where energy conditions were satisfied.


 The motivation of this paper is to develop a new shape function to  study the wormhole solutions in different theories of gravitation. Therefore, in this paper, a new shape function is defined and wormhole solutions are explored in (i) $f(R)$ gravity, with specific form of $f(R)=R-\mu R_c\tanh \frac{R}{R_c}$,
  where $R$ is scalar curvature, $\mu$ and $R_c$ are positive constants, (ii) $f(R,T)$ gravity, with specific form of $f(R, T)=R+2f(T)$, where $f(T)=\lambda T$, $T$ is the trace of the energy-momentum tensor and $\lambda$ is a constant and (iii) general relativity. The section-wise description is as follows:
in Section 2, the wormhole structure is discussed. In Sections 3 and 4,  brief reviews on $f(R)$ and $f(R,T)$ gravity theories, respectively, are presented. In Section 5, the field equations are solved and energy condition terms are computed, plotted and analyzed with respect to both modified theories. In Section 6, the findings are compared and finally, in Section 7, the work is concluded.

\section{Wormhole Structure}
A static and spherically symmetric wormhole structure is defined by the metric
\begin{equation}\label{metric}
ds^2=-e^{2\Phi(r)}dt^2+\frac{dr^2}{1-b(r)/r} + r^2(d\theta^2+\sin^2\theta d\phi^2),
\end{equation}
where the functions $b(r)$ and $\Phi(r)$ are called as shape and redshift functions, respectively.
The radial coordinate $r$ varies from $r_0$ to $\infty$, where $r_0$ is called the radius of the throat. The angles $\theta$ and $\phi$ vary from 0 to $\pi$ and 0 to $2\pi$, respectively. To avoid the presence of horizons and singularities, the redshift function should be finite and non-zero.  The shape function should satisfy the following properties: (i) $\frac{b(r)}{r}<1$ for $r>r_0$,  (ii) $b(r_0)=r_0$ at $r=r_0$, (iii) $\frac{b(r)}{r}\rightarrow 0$ as $r\rightarrow \infty$, (iv) $\frac{b(r)-b'(r)r}{b(r)^2}>0$ for $r>r_0$ and (v) $b'(r_0)\leq1$. The condition (i) is necessary for the radial metric component to be negative. The shape function possesses minimum value equal to $r_0$, given by condition (ii). To obtain asymptotically flat space time as $r\rightarrow \infty$, the condition (iii) is required. Conditions (iv) and (v) are known as flaring out conditions, which are required to obtain traversable wormholes.

Lobo and Oliveira \cite{Lobo1} chosen some specific shape functions $(i)$ $b(r) = \frac{r_0^2}{r}$ and $(ii)$ $b(r) =\sqrt{r_0r}$ to study traversable wormholes in $f(R)$ gravity. Rahaman et al. \cite{Rahaman} considered shape functions  $(i)$ $b(r) = r_0+\rho_0r_0^3\ln(\frac{r_0}{r})$ and $(ii)$ $b(r) = r_0+\gamma r_0(1-\frac{r_0}{r})$, where $\rho_0$ and $\gamma$ were arbitrary constants and less than unity, to explore Finslerian wormhole models.  Cataldo et al. \cite{cataldo1} assumed a linear shape function $b(r)=\alpha + \beta r$, where $\alpha$ and $\beta$ were arbitrary constants and obtained wormholes connecting two asymptotic non-flat regions with a solid angle deficit. Jahromi and Moradpour \cite{moradpour} considered shape function $b(r) = a \tanh(r)$, where $a$ was a constant and explored static traversable wormholes in Lyra geometry.
Kuhfittig \cite{peter} took the shape function $b(r)=r_0(\frac{r}{r_0})^\beta$, where $0<\beta<1$, to investigate the existence of traversable wormholes in the background of $f(R)$ gravity. Recently, Godani \& Samanta \cite{ns}  defined a shape function $b(r)=\frac{r_0\log(r+1)}{\log(r_0+1)}$ and studied energy conditions in the context of traversable wormholes. Since the shape functions have important role for wormhole modeling, in this paper, we have defined a new shape function $b(r)$ which is as follows:
\begin{equation}\label{shape}
b(r)=\frac{r}{\exp[(r-r_0)/L]},
\end{equation}
where ``$L$" is a characteristic length scale of the wormhole throat. 
\begin{figure}
\begin{center}
		\includegraphics[height=10cm, width=14cm]{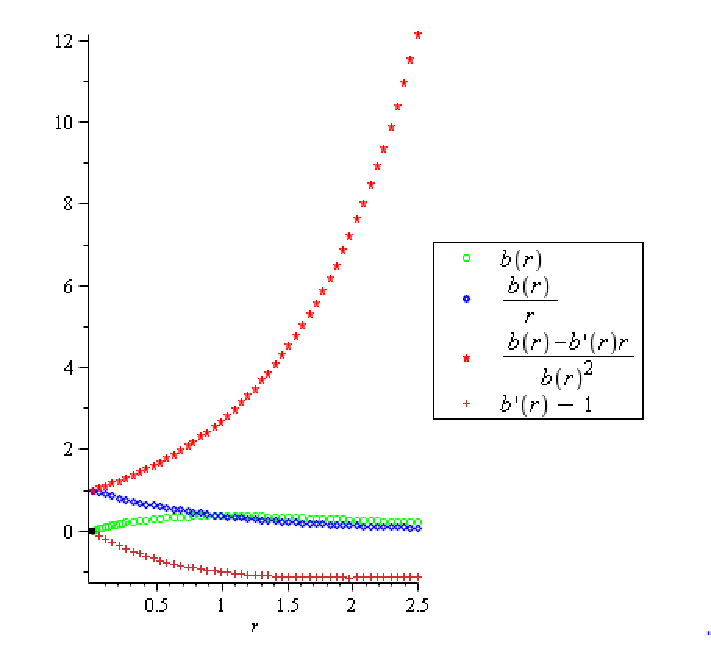}
	\caption{Plot for various conditions satisfied by the shape function $b(r)$}
\end{center}

\end{figure}
\\

\noindent
There is no physical reason or underlying theory for the shape function \eqref{shape}. This shape function satisfies all the conditions discussed above as shown in Fig. (1). Now, we have explored the significance of this shape function in two directions. Firstly, we have used the embedding diagram to impose the demand that the wormhole is described by the metric (1).  Secondly, we have determined radial and lateral tidal constraints for Zero-tidal-force-Schwarzschild-like wormholes and consequently, calculated an upper bound for the speed of traveler at $r=r_0$.
We have considered spherically symmetric geometry for equatorial slice $\theta=\frac{\pi}{2}$ at a fixed time. Using $\theta=\frac{\pi}{2}$ and $t=$ constant, the metric \eqref{metric} takes the form
\begin{equation} \label{tcons}
ds^2=\frac{1}{1-\frac{b(r)}{r}}dr^2+r^2d\phi^2.
\end{equation}

\noindent
We are interested to  envisage this slice as taken away from 4-dimensional space-time and embedded in 3-dimensional Euclidean space as a 2-dimensional surface having the geometry same as the slice considered above.
In cylindrical coordinates, the metric for the 3-dimensional Euclidean space is given by
\begin{equation} \label{cyl}
ds^2=dr^2+r^2d\phi^2+dz^2.
\end{equation}
The embedded surface can be described by $z\equiv z(r)$ so that it will be axially symmetric. Therefore, the line element \eqref{cyl} can be written as
\begin{equation} \label{cyl1}
ds^2=\Big(1+\Big(\frac{dz}{dr}\Big)^2\Big)dr^2+r^2d\phi^2.
\end{equation}
 Identifying the coordinates $(r, \phi)$ of the embedding space with the coordinates $(r, \phi)$ of the wormhole space-time, from Eqns. \eqref{tcons} \& \eqref{cyl1}, we have
\begin{equation}\label{zr}
\frac{dz}{dr}=\pm\Big(\frac{r}{b(r)}-1\Big)^{-1/2}.
\end{equation}
The solution of Eq. (\ref{zr}) gives the embedded surface which is plotted in Fig. (2).
At $r=r_0$, this embedding surface is ill defined, i.e. $\frac{dz}{dr}=\infty$ at $r=r_0$. The outside space which is far from the throat of wormhole is asymptotically flat.

\begin{figure}
	\centering
	\includegraphics[scale=.5]{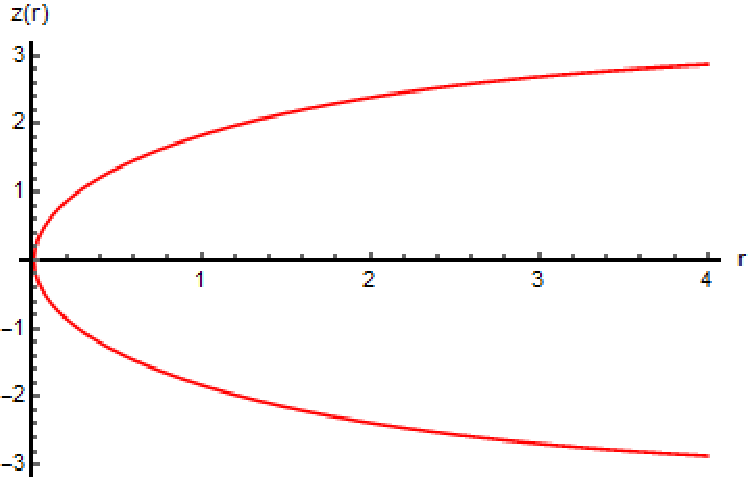}
	\caption{Plot of embedding surface $z(r)$ with respect to $r$ for slice $t$ = constant and $\theta=\pi/2$}
\end{figure}

\noindent
Now, the proper radial distance is defined as \cite{morris1}
\begin{eqnarray}\label{}
l(r)&=&\pm\int_{r_0}^{r}\left(\frac{r-b(r)}{r}\right)^{-\frac{1}{2}}dr.
\end{eqnarray}
Using shape function \eqref{shape},
\begin{eqnarray}\label{}
l(r)&=&\pm\int_{r_0}^{r}\left(1-\exp(r_0-r)\right)^{-\frac{1}{2}}dr.
\end{eqnarray}
For the upper universe of wormhole, $z>0$ and hence $l(r)$ is positive. Similarly, for the lower universe, $z<0$ and hence $l(r)<0$. For proper radial distance $l(r)$ to be well behaved everywhere, it should be finite everywhere in space-time. At a far distance from the throat of wormhole, the space becomes asymptotically flat, so $\frac{dz}{dr}\rightarrow 0$ as $l(r)\rightarrow \pm \infty$.

%
We are studying traversable wormhole, therefore it should not contain any horizon. A horizon, if present, would prevent two way travel through wormhole. Hence, to maintain traversability condition, $\Phi(r)$ must be finite. Therefore, in this study, we assume
$\Phi(r)\equiv \mbox{constant}$, so that $\Phi^{'}\equiv 0$. The condition $\Phi^{'}\equiv 0$, called the zero-tidal-force solution in \cite{morris1}, is highly desirable
feature for a traversable condition in wormhole.
Zero-tidal-force-Schwarzschild-like wormholes form a specific and simple class of wormholes for which the metric \eqref{metric} takes the form
\begin{equation}\label{metric2}
ds^2=-dt^2+\frac{dr^2}{1-b(r)/r} + r^2(d\theta^2+\sin^2\theta d\phi^2).
\end{equation}
Now, we focus on the traversability condition of such wormholes. Suppose a traveler starts journey radially from a point in the lower universe to a point in the upper universe. Then the acceleration experienced by the traveler should not be greater than $g=9.8$ m/sec$^2$.
In \cite{morris1, morris2}, the acceleration experienced by a traveler traveling radially is defined as
\begin{eqnarray}
a&=&\pm \sqrt{1-\frac{b(r)}{r}}\exp^{-\phi(r)}(\gamma\exp{\phi(r)})'c^2\nonumber\\
&=&\pm \sqrt{1-\exp(r_0-r)}\exp^{-\phi(r)}(\gamma\exp{\phi(r)})'c^2,
\end{eqnarray}
where $\gamma=(1-v^2/c^2)^{-1}$, $v$ is the radial velocity of traveler and $c$ is the speed of light.
Then for the zero-tidal-force-Schwarzschild-like wormholes (\ref{metric2}), the constraint is given by\\
$|a|=|\sqrt{1-\exp(r_0-r)}\gamma^{'}c^2|\leq g$.

Another traversability condition is associated with
tidal accelerations felt by the traveler that should not be greater than the Earth's gravitational acceleration. In \cite{morris1}, radial and lateral tidal constraints are defined. The radial tidal constraint provides a constraint on metric coefficient $\phi$ and the lateral tidal constraint provides a constraint on the speed of the traveler with which it passes the wormhole. For a traveler, having the size of the body equal to $\epsilon$, the expressions for radial and lateral tidal constraints, respectively, are given as
\begin{equation}\label{cons1}
\Big|(1-\frac{b}{r})(-\phi^{''}-\phi^{'2}+\frac{b'r-b}{2r(r-b)}\phi')c^2\Big||\epsilon|\leq g
\end{equation} and
\begin{equation}\label{cons2}
\Big|\frac{\gamma^2c^2}{2r^2}\Big[\frac{v^2}{c^2}(b'-\frac{b}{r})+2(r-b)\phi^{'}\Big]\Big||\epsilon\|\leq g.
\end{equation}

\noindent
For metric \eqref{metric2}, the lateral tidal constraint gives $\Big|-\exp(r_0-r)\frac{\gamma^2v^2}{2r}\Big||\epsilon|\leq g$, however the radial tidal constraint vanishes everywhere. For the non-relativistic motion, $v<<c$ which implies that $\gamma\approx 1.$ Thus, at $r=r_0$,   an upper bound for traverler's speed is obtained as $v\leq\sqrt{9.8r_0}.$


\section{Field Equations in $f(R)$ Gravity}
The  gravitational action for Einstein's theory of general relativity is defined as
\begin{equation}\label{action}
S_G=\dfrac{1}{16\pi}\int[R + L_m]\sqrt{-g}d^4x,
\end{equation}
where $R$ is the scalar curvature, $L_m$ is the  matter Lagrangian density and $g$ is the  determinant of the metric $g_{\mu\nu}$.
It was found inadequate to describe the present accelerating phase of our universe.  Consequently, the action \eqref{action} was generalized by replacing  $R$ with an arbitrary function $f(R)$. It is defined as
\begin{equation}\label{action1}
S_G=\dfrac{1}{16\pi}\int[f(R) + L_m]\sqrt{-g}d^4x.
\end{equation}

\noindent
Variation of Eq.(\ref{action1}) with respect to the metric $g_{\mu\nu}$ gives the field equations as
\begin{equation}\label{fe}
FR_{\mu\nu} -\dfrac{1}{2}fg_{\mu\nu}-\triangledown_\mu\triangledown_\nu F+\square Fg_{\mu\nu}= T_{\mu\nu}^m,
\end{equation}	
where $R_{\mu\nu}$ denotes Ricci tensor and $F=\frac{df}{dR}$. The contraction of \ref{fe}, gives
\begin{equation}\label{trace}
FR-2f+3\square F=T,
\end{equation}
where $T=T^{\mu}_{\mu}$ is the trace of the stress energy tensor.
From Eqs. \ref{fe} \& \ref{trace}, the effective field equations are obtained as
\begin{equation}
G_{\mu\nu}\equiv R_{\mu\nu}-\frac{1}{2}Rg_{\mu\nu}=T_{\mu\nu}^{eff},
\end{equation}
where $T_{\mu\nu}^{eff}=T_{\mu\nu}^{c}+T_{\mu\nu}^{m}/F$ and $T_{\mu\nu}^{c}=\frac{1}{F}[\triangledown_\mu\triangledown_\nu F-\frac{1}{4}g_{\mu\nu}(FR+\square F+T)]$.
The energy momentum tensor for the matter source of the wormhole is $T_{\mu\nu}=\frac{\partial L_m}{\partial g^{\mu\nu}}$, which is defined as
\begin{equation}
T_{\mu\nu} = (\rho + p_t)u_\mu u_\nu - p_tg_{\mu\nu}+(p_r-p_t)X_\mu X_\nu,
\end{equation}	
such that
\begin{equation}
u^{\mu}u_\mu=-1 \mbox{ and } X^{\mu}X_\mu=1,
\end{equation}

\noindent
where $\rho$,  $p_t$ and $p_r$  stand for the energy density, tangential pressure and radial pressure, respectively.
The  Ricci scalar $R$ is given by $R=\frac{2b'(r)}{r^2}$ and Einstein's field equations for the metric \ref{metric} in  $f(R)$ gravity are obtained as:
\begin{equation}\label{6}
\rho=\frac{Fb'(r)}{r^2}-H
\end{equation}
\begin{equation}\label{7}
p_r=-\frac{b(r)F}{r^3}-\Bigg(1-\frac{b(r)}{r}\Bigg)\Bigg[F''+\frac{F'(rb'(r)-b(r))}{2r^2\Big(1-\frac{b(r)}{r}\Big)}\Bigg]+H
\end{equation}
\begin{equation}\label{8}
p_t=\frac{F(b(r)-rb'(r))}{2r^3}-\frac{F'}{r}\Bigg(1-\frac{b(r)}{r}\Bigg)+H,
\end{equation}

\noindent
where $H=\frac{1}{4}(FR+\square F+T)$ and prime upon a function denotes the derivative of that function with respect to  radial coordinate $r$.

%
%
%
%
%
%
%
%

\section{Field Equations in $f(R, T)$ Gravity}
Harko et al. \cite{harko} modified Einstein's general relativity by replacing $R$ with an arbitrary function $f(R,T)$ of $R$ and $T$, where $T$ is the trace of the energy-momentum tensor  and defined the gravitational  action as
\begin{equation}\label{action2}
S_G=\dfrac{1}{16\pi}\int[f(R,T)+L_m]\sqrt{-g}d^4x.
\end{equation}

\noindent
Let $\square \equiv- \triangledown^{\mu}\triangledown_{\nu}$ and
$\theta_{\mu\nu} = -2T_{\mu\nu} + g^{\mu\nu}L_m - 2 g^{\gamma\sigma}\dfrac{\partial^2 L_m}{\partial g^{\mu\nu}\partial g^{\gamma\sigma}}$.

\noindent
Taking $L_m=-\rho$,

\begin{equation}\label{theta}
\theta_{\mu\nu} = -2T_{\mu\nu} - \rho g^{\mu\nu}.
\end{equation}

%
\noindent
Varying action (\ref{action2}) with respect to the metric, field equations are
\begin{equation}\label{frt}
f(R,T)R_{\mu\nu} -\frac{1}{2}f(R,T)g_{\mu\nu} + (g_{\mu\nu}
\square
-\triangledown_\mu\triangledown_\nu)f_R(R,T)=8\pi T_{\mu\nu} - f_T(R,T) \theta_{\mu\nu},
\end{equation}
where $f_R(R,T) \equiv \dfrac{\partial f(R, T)}{\partial R}$ and
$ f_T(R,T) \equiv \dfrac{\partial f(R,T)}{\partial T}.$

\noindent
For $f(R,T)=R + 2f(T)$, where $T$ is the trace of the energy-momentum tensor, the gravitational field equations  from Eq.\eqref{frt} are obtained as
\begin{equation}\label{efe}
R_{\mu\nu}-\frac{1}{2}Rg_{\mu\nu} = 8\pi T_{\mu\nu} - 2f^{'}(T)T_{\mu\nu}-2f^{'}(T)\theta_{\mu\nu}+f(T)g_{\mu\nu},
\end{equation}
where $'$ stands for the derivative of $f(T)$ with respect to $T$.
Using Eq.\eqref{theta} in Eq.\eqref{efe},

\begin{equation}\label{efe1}
R_{\mu\nu}-\frac{1}{2}Rg_{\mu\nu} = 8\pi T_{\mu\nu} + 2f^{'}(T)T_{\mu\nu}+[2\rho f^{'}(T)+f(T)]g_{\mu\nu}.
\end{equation}

\noindent
For $f(T)=\lambda T$, where $\lambda$ is a constant,
the field equations for the wormhole metric \eqref{metric} are come out to be
\begin{equation}\label{frt1}
  \frac{b^{'}}{r^2}=(8\pi+\lambda)\rho-\lambda(p_r+2p_t),
\end{equation}

\begin{equation}\label{frt2}
  -\frac{b}{r^3}=\lambda \rho+(8\pi+3\lambda)p_r+2\lambda p_t,
\end{equation}

\begin{equation}\label{frt3}
  \frac{b-b^{'}r}{2r^3}=\lambda \rho+\lambda p_r+(8\pi+4\lambda)p_t.
\end{equation}

%
%

\section{Wormhole Solutions}

\noindent
Tsujikawa \cite{Tsujikawa} introduced a viable cosmological $f(R)$ model with the function $f(R)$ defined as
\begin{equation}
f(R)=R-\mu R_c\tanh \frac{R}{R_c},
\end{equation}
 where $\mu$ and $R_c$  are positive constants. The  model is cosmologically viable and  satisfies  local gravity constraint.
Harko et al. \cite{harko} proposed an $f(R,T)$ model with the function $f(R,T)=R+2\lambda T$, where $\lambda$ is a constant and $T$ is the trace of energy-momentum tensor.

In this section, the solutions of wormhole metric \eqref{metric} and energy condition terms are derived and plotted with respect to above $f(R)$ and $f(R,T)$ models. The conclusions drawn are also analyzed for each set of plots.

\subsection{Case 1: For $f(R)$ model}
\noindent
The explicit form of energy density, radial pressure, tangential pressure and all  energy conditions terms are presented in Appendix A (please see
the derivations in Appendix A (Eqns. (46)-(54))). 
\\

\noindent
The anisotropy parameter ($\triangle$) is defined as $\triangle=p_t-p_r$. If $\triangle<0$, then the geometry is said to be attractive; if $\triangle>0$, then the geometry is said to be repulsive and if $\triangle = 0$ , then the geometry has an isotropic pressure.  The equation of state parameter ($\omega$) in terms of radial pressure and density is defined as $w = \frac{p_r}{\rho}$. It is also called as radial state parameter.
\begin{figure}
	\centering
	\subfigure[In this figure, $\log_{10}(\rho+p_r)$ is plotted with respect to $r$. The NEC term $\rho + p_r$ is  positive for $58.75<r<63.72$, which is shown here using logarithmic scale on vertical axis and $L=1$. ]{\includegraphics[scale=.95]{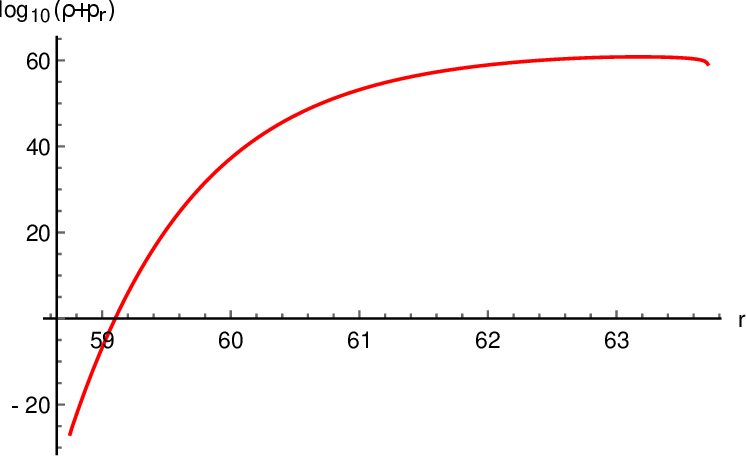}}
	\vspace{1cm}
	\subfigure[In this figure, $\log_{10}(\rho+p_t)$ 
	is plotted with respect to $r$. The NEC term $\rho + p_t$ is  positive for
	$0<r\leq2$, $59.25<r<\infty$ and $r\neq 1$, which is shown here using logarithmic scale on vertical axis and $L=1$.
	]{\includegraphics[scale=.95]{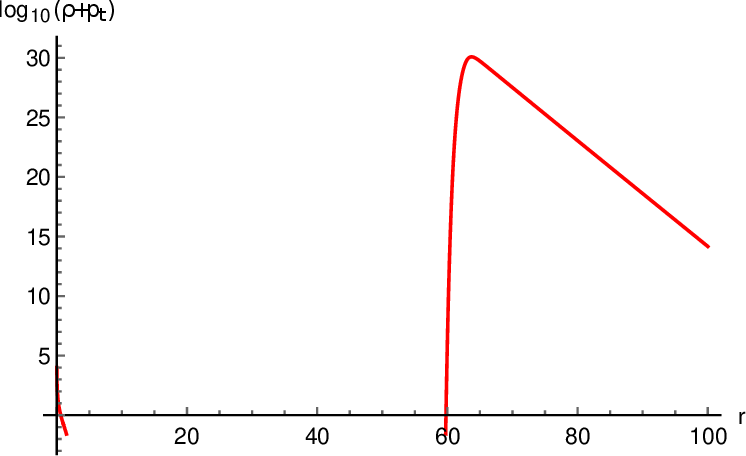}}
	\caption{Plots for Null Energy Condition (NEC) terms in $f(R)$ gravity}
\end{figure}
\\

  \noindent
  The behaviour of the energy density is  positive  for $r\in(0,1)$, zero for $r=1$ and negative for $r\in(1,\infty)$. This indicates  the presence of exotic matter for $r\in(1,\infty)$.\\

  \noindent
  The null energy condition (NEC) in terms of pressures is defined as $\rho+p_r \geq 0$ and $\rho+p_t \geq 0$. 
  The first NEC term $\rho+p_r$ is negative for $r\in [44.61,58.75]\cup[63.72,\infty)$,   positive for $r\in (58.75,63.72)$. For rest values of $r$, i.e. for $r\in (0,44.61)$, $\rho+p_r$ has indeterminate value. In Fig. 3(a), it is plotted for $r\in (58.75,63.72)$ using logarithmic scale on vertical axis. \\

  \noindent
  The second NEC term, i.e. $\rho+p_t$ is plotted with respect to $r$ in Fig. 3(b). It is  positive for $r\in(0,2]\cup(59.25,\infty)-\{1\}$ and negative for $r \in\{1\}\cup (2, 59.25]$. In Fig. 3(b), it is plotted for $r\in (58.75,63.72)$ using logarithmic scale on vertical axis.
  Hence, both  NEC terms are positive  only for $r\in (59.25,63.72)$.  Therefore, NEC is satisfied for $r\in (59.25,63.72)$ and  violated elsewhere.\\

\noindent
  The weak energy condition (WEC) in terms of pressures is defined as $\rho\geq 0$, $\rho+p_r \geq 0$ and $\rho+p_t \geq 0$.
 Since $\rho\geq 0$ for $r\in (0,1]$ and NEC is satisfied for  $r\in (59.25,63.72)$,  WEC is violated everywhere.
\\

\noindent
The dominant energy condition (DEC) is defined as $\rho-|p_r|\geq 0$, $\rho-|p_t|\geq 0$ and $\rho\geq 0$.
The first DEC term is obtained to have indeterminate value for $r\in(0,44.71)$ and negative values for $r\in [44.71,\infty)$.
The second DEC term is positive for $r\in (0,0.67]$ and negative for $r\in(0.67,\infty)$.
Subsequently, DEC is violated throughout.
\\

\noindent
  The anisotropy parameter $\triangle=p_t-p_r$ is
  found to have indeterminate value for $r\in(0,44.71)$ and positive values for $r\in[44.71,58.84)$.
  Consequently, it is negative for $r\in[58.84,63.81]$ and positive for $r\in (63.81,\infty)$. This indicates that there is  repulsive geometry for $r\in[44.71,58.84)\cup(63.81,\infty)$, attractive geometry for $r\in[58.84,63.81]$ and there is no idea about the nature of the geometry (i.e. whether the geometry is repulsive or attractive) for $r\in(0,44.71)$.
  This shows that the geometry is partly  attractive and repulsive.
\\

\noindent
   The equation of state parameter $\omega$
   has indeterminate value for $r\in(0,44.71)$, $0<\omega< 1$ for $r\in [44.71,58.83)$ and  $\omega<-1$ for $r\in [58.83,\infty)$.
   This indicates that  there is no idea about the matter contained in the wormholes for $r\in(0,44.71)$,  the wormholes may contain an ordinary type of matter for $r\in [44.71,58.83)$ and  the wormholes contain phantom fluid for $r\in [58.83,\infty)$.  This shows that the wormholes are not filled with same type of matter for all values of $r$.

\subsection{Case 2: For $f(R,T)$ model}

\noindent
From Equations \eqref{frt1}, \eqref{frt2}, \eqref{frt3} and \eqref{shape}, we have

%
\begin{eqnarray}
	\rho&=&\frac{(L-r) e^{\frac{r_0-r}{L}}}{2 (\lambda +4 \pi ) L r^2}\label{r1}\\
	p_r&=&-\frac{e^{\frac{r_0-r}{L}}}{2 (\lambda +4 \pi ) r^2} \label{pr1}\\
	p_t&=&\frac{e^{\frac{r_0-r}{L}}}{4Lr(\lambda  +4 \pi)} \label{pt1}
\end{eqnarray}

\noindent
From Equations \eqref{r1}, \eqref{pr1} and \eqref{pt1},
\begin{eqnarray}
	\rho+p_r&=&-\frac{e^{\frac{r_0-r}{L}}}{2 (\lambda +4 \pi ) L r} \\
	\rho+p_t&=&\frac{(2 L-r) e^{\frac{r_0-r}{L}}}{4 (\lambda +4 \pi ) L r^2} \\
	\rho-|p_r|&=&\frac{(L-r) e^{\frac{r_0-r}{L}}}{2 (\lambda +4 \pi ) L r^2}-\Big|-\frac{e^{\frac{r_0-r}{L}}}{2 (\lambda +4 \pi ) r^2}\Big| \\
	\rho-|p_t|&=&\frac{(L-r) e^{\frac{r_0-r}{L}}}{2 (\lambda +4 \pi ) L r^2}-\Big|\frac{e^{\frac{r_0}{L}-\frac{r}{L}}}{4 \lambda  L r+16 \pi  L r} \Big|\label{pt}\\
	\end{eqnarray}

\begin{eqnarray}
p_t-p_r&=&\frac{(r+2 L) e^{\frac{r_0-r}{L}}}{4 (\lambda +4 \pi ) L r^2}\\
\frac{p_r}{\rho}&=&-\frac{L}{L-r}\label{w}
\end{eqnarray}

\begin{figure}{}
	\centering
	\subfigure[The NEC term $\rho + p_r$ is positive for $r\leq2$ with $\lambda>-4\pi$. In this figure, $\rho + p_r$ is plotted for $\lambda=-\pi, 10 \pi$ and $20 \pi$ with  $L=1$. ]{\includegraphics[scale=.41]{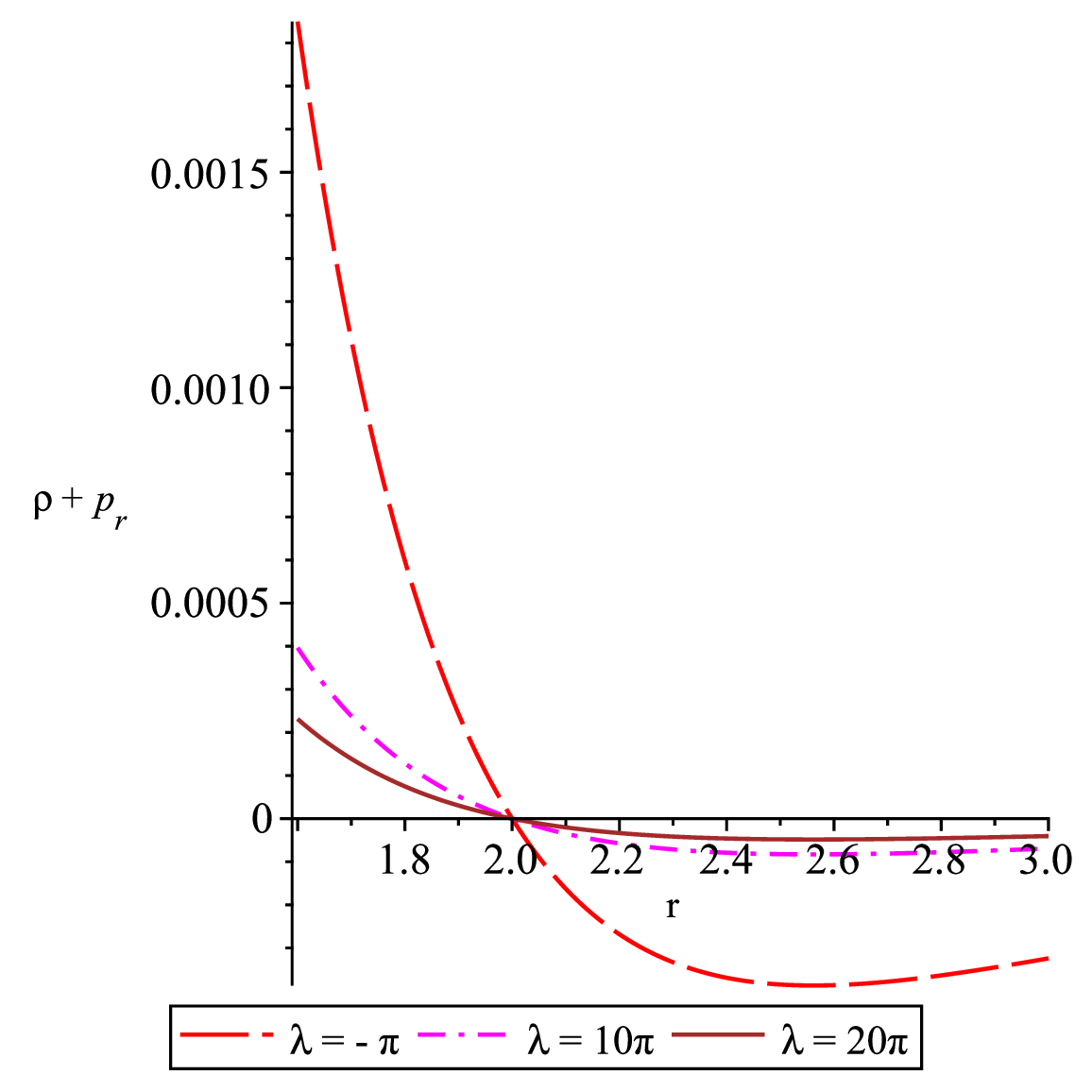}}\hspace{.65cm}
	\subfigure[The NEC term is positive $\rho + p_r$ for $r\geq2$ with $\lambda<-4\pi$. In this figure, $\rho + p_r$ is plotted for $\lambda=-10\pi, -20 \pi$ and $-30 \pi$ with  $L=1$. ]{\includegraphics[scale=.41]{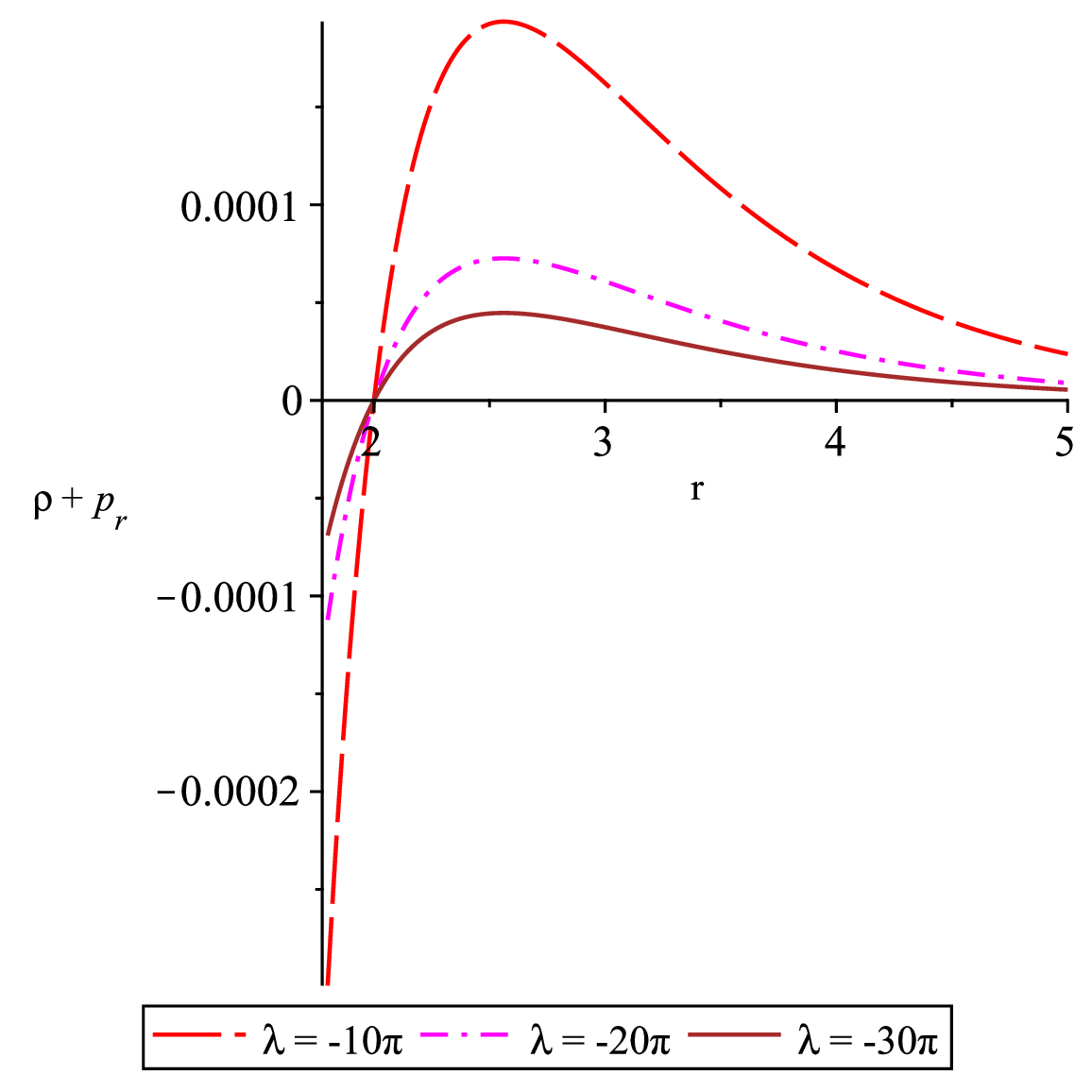}}
	\subfigure[The NEC term $\rho + p_t$ is positive for  $r\leq2$ with $\lambda>-4\pi$. In this figure, $\rho + p_t$ is plotted for $\lambda=-10\pi, -20 \pi$ and $-30 \pi$ with  $L=1$.]{\includegraphics[scale=.42]{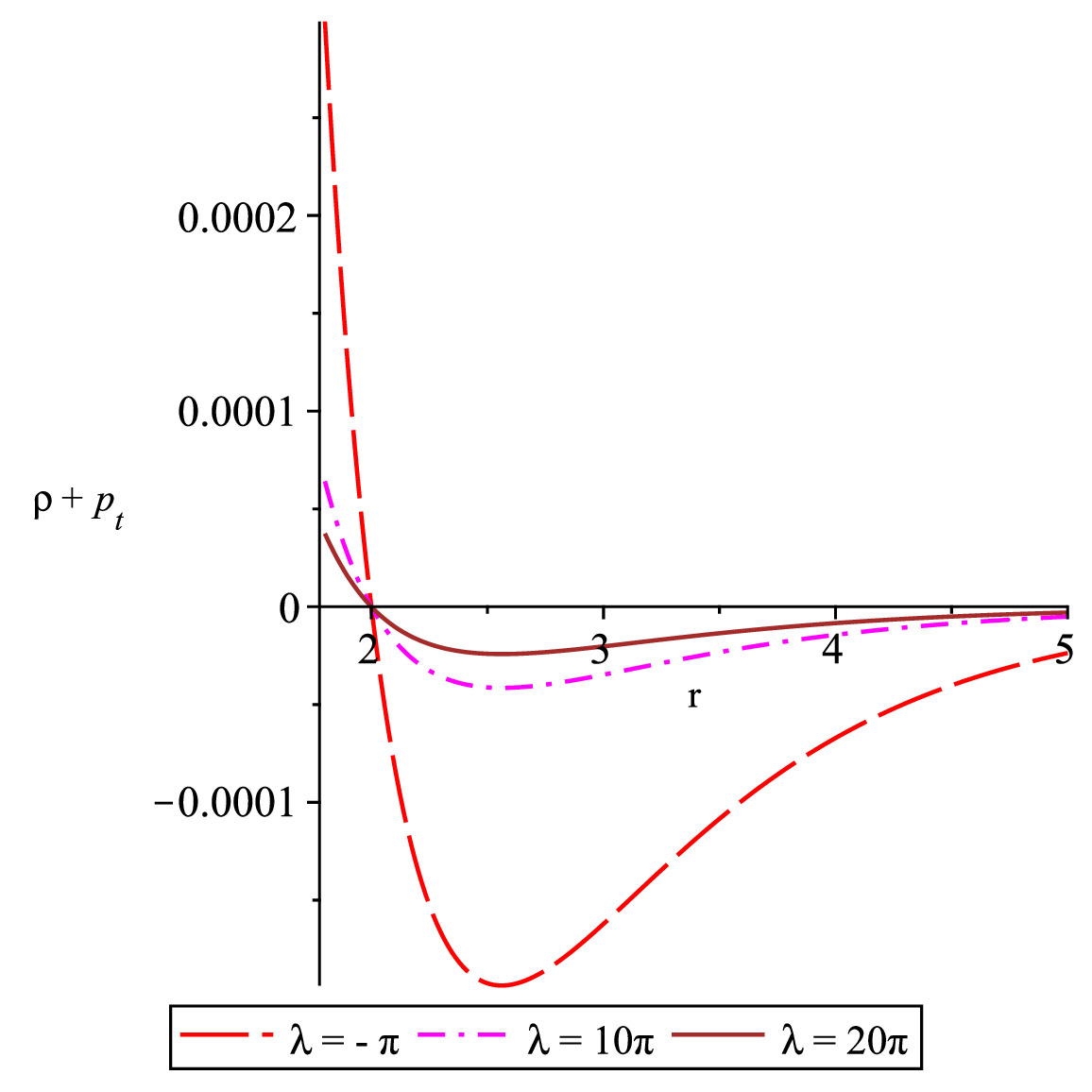}}\hspace{.65cm}
	\subfigure[The NEC term $\rho + p_t$ is positive for  $r\geq2$ with $\lambda<-4\pi$. In this figure, $\rho + p_t$ is plotted for $\lambda=-\pi, 10 \pi$ and $20 \pi$ with  $L=1$.]{\includegraphics[scale=.42]{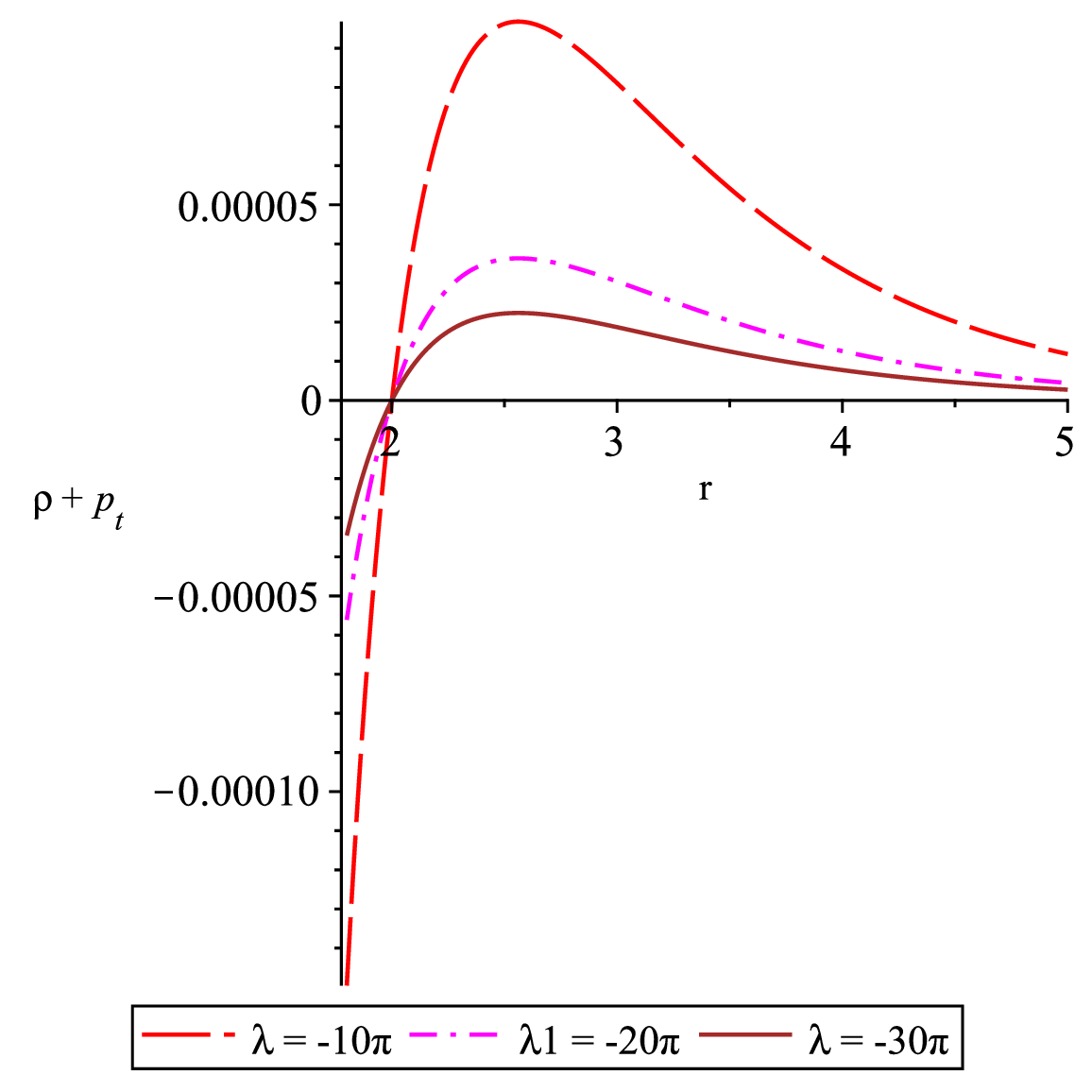}}
	\caption{These are plots for null energy condition (NEC) in $f(R,T)$ gravity, which is satisfied for (i) $r<2$ with $\lambda>-4\pi$ and (ii) $r\geq2$ with $\lambda<-4\pi$. From the above plots, it is observed that if we take $\lambda<-4\pi$, then NEC  is satisfied for $r\geq 2$ and violated for $r<2$. Hence, $\lambda<-4\pi$ is a suitable choice of parameter $\lambda$ to obtain the wormhole geometry filled with less amount of exotic matter, in $f(R, T)$ gravity case.}
\end{figure}

\noindent
In Equations \eqref{r1}-\eqref{w}, the expressions for the energy density, radial and tangential pressures and their different combinations are computed in the context of $f(R,T)$ gravity. The behaviour of energy density, radial and tangential pressures, NEC and DEC terms, anisotropy parameter and equation of state parameter are studied in detail. 

The nature of the energy density ($\rho$) is  positive, i.e. $\rho\geq 0$  for (i) $r\in(0,1]$ with $\lambda>-4\pi$ and (ii) $r\in[1,\infty)$ with $\lambda<-4\pi$. For other combinations of ranges for $r$ and $\lambda$,  it is  negative.
Hence, it is observed that the entire wormhole does not contain either positive energy density or negative energy density.

\noindent
The first NEC term $\rho+p_r$ is positive for (i) $r\in(0,2]$ with $\lambda>-4\pi$ (ii)  $r\in[2,\infty)$ with $\lambda<-4\pi$. It is negative for (i) $r\in(2,\infty)$ with $\lambda>-4\pi$ (ii)  $r\in(0,2]$ with $\lambda<-4\pi$.  
In Figs. 4(a) and 4(b),  the first NEC term $\rho+p_r$ is plotted. In Fig. 4(a), $\rho+p_r$ is shown to be positive for $r\in(0,2]$ by taking $\lambda=-\pi, 10\pi$ and $20\pi$ and in Fig. 4(b), $\rho+p_r$ is shown to be positive for $r\in[2,\infty)$ by taking $\lambda=-10\pi, -20\pi$ and $-30\pi$.\\

\noindent
The second NEC term $\rho+p_t$ is also positive for (i) $r\in(0,2]$ with $\lambda>-4\pi$ (ii)  $r\in[2,\infty)$ with $\lambda<-4\pi$. It is negative for (i) $r\in(2,\infty)$ with $\lambda>-4\pi$ (ii)  $r\in(0,2]$ with $\lambda<-4\pi$.  
In Figs. 4(c) and 4(d),  the second NEC term $\rho+p_t$ is plotted. In Fig. 4(c), $\rho+p_t$ is shown to be positive for $r\in(0,2]$ by taking $\lambda=-\pi, 10\pi$ and $20\pi$ and in Fig. 4(d), $\rho+p_t$ is shown to be positive for $r\in[2,\infty)$ by taking $\lambda=-10\pi, -20\pi$ and $-30\pi$.\\

\noindent
Thus, both NEC terms are positive for (i) $r\in(0,2]$ with $\lambda>-4\pi$ and (ii) $r\in[2,\infty)$ with $\lambda<-4\pi$. Since either $\lambda>-4\pi$ or $\lambda<-4\pi$.
If we consider $\lambda>-4\pi$, then NEC  is satisfied for $r\in(0,2]$ and violated for $r>2$. If we take $\lambda<-4\pi$, then NEC  is satisfied for $r\in[2,\infty)$ and violated for $r<2$. For the rest combinations of ranges for $r$ and $\lambda$, NEC is violated.
Thus, we have obtained a wide range of $r$ for the satisfaction of NEC with $\lambda<-4\pi$.\\

\noindent
From the results discussed above for the energy density and NEC, we find that WEC is also satisfied for (i) $r\in(0,1]$ with  $\lambda>-4\pi$ and (ii) $r\in[2,\infty)$ with  $\lambda<-4\pi$.
\\

\noindent
The first DEC term is negative for $r\in(0,2]$ with $\lambda>-4\pi$ and  positive for $r\in[2,\infty)$ with $\lambda<-4\pi$.
The second DEC term is positive for (i) $r\in(0,2]$ with $\lambda>-4\pi$ and (ii) $r\in[2,\infty)$ with $\lambda<-4\pi$.
Thus, DEC is violated for $r\in(0,2]$ with $\lambda>-4\pi$ and satisfied for $r\in[2,\infty)$ with $\lambda<-4\pi$.
\\

\noindent
The anisotropic parameter ($\triangle$) is   negative for (i) $r\in(0,2)$ with $\lambda>-4\pi$ and (ii) $r\in(2,\infty)$ with $\lambda<-4\pi$, which represents an attractive nature of geometry in these intervals. Further, $\triangle$ is found to possess zero value at $r=2$ with $\lambda\neq -4\pi$, which indicates presence of isotropic pressures, i.e. equality of $p_r$ and $p_t$.

\noindent
The equation of state parameter is a positively increasing function for $r\in(0,1)$. For $r=1$, it possesses infinite value. Moreover, it is a negatively increasing function for $r\in(1,\infty)$. For $r\in(1,2)$, $\omega<-1$; for $r=2$, $\omega=-1$ and for $r\in(2,\infty)$, $-1<\omega<0$.


Hence, for $r\in[2,\infty)$, where energy conditions are  satisfied with $\lambda<-4\pi$, the  wormholes are filled with quintessence or non-phantom fluid.
Logically, we should consider either $\lambda>-4\pi$ or $\lambda<-4\pi$, because at a time two ranges can not be possible. Hence, we can restrict for $\lambda<-4\pi$ because for this range of $\lambda$, we have got a large range of $r$ for the satisfaction of NEC, WEC and DEC.

Taking $\lambda=0$, the $f(R,T)$ model reduces to GR. For this particular case, it is observed that for $r\in(0,1)$, $\rho>0$; for $r=1$, $\rho=0$ and for $r\in(1,\infty)$, $\rho<0$.  Both null energy condition terms $\rho+p_t$ and $\rho+p_r$ are positive for $r\in(0,2)$, zero for $r=2$ and negative for $r\in(2,\infty)$. This shows the satisfaction of NEC for $r\in(0,2]$ and its dissatisfaction for $r\in(2,\infty)$. Consequently, WEC is satisfied for $r\in(0,1]$. The first DEC term $\rho-|p_r|$ is  negative for all $r\in(0,\infty)$. However, the second DEC term  $\rho-|p_t|$ is positive for $r\in(0, 0.66]$ and negative for $r\in(0.66,\infty)$. This shows the violation of  DEC everywhere. The anisotropy parameter is  negative for $r\in(0,2)$, zero for $r=2$ and positive for $r\in(2,\infty)$. This indicates that the geometry is attractive near the throat and repulsive outside the throat. The equation of state parameter for $f(R,T)$ model is  independent of $\lambda$, therefore, the results are same as discussed above in case of $f(R,T)$.

\section{Results \& Discussion}

\noindent
The alternate theories have  significant role in the exploration of wormhole geometries. The metric of wormhole depends on two functions: shape function and redshift function. For simplicity and to avoid horizon, the redshift function is taken constant. However, the shape function that contributes for the shape of the wormhole is newly defined  in terms of exponential function as $b(r)=\frac{r}{\exp[(r-r_0)/L]}$. Using this shape function, the existence of wormhole solutions is investigated by determining  various energy conditions, anisotropy parameter and equation of state parameter  in the context of $f(R)$, $f(R,T)$ and general relativity theories.  The results obtained are summarized in Tables 1-3. With the help of these tables, the results are compared below:
\begin{table}[!h]
	\centering
	\caption{Summary of results in $f(R)$ gravity}
	\begin{tabular}{|c|c|l|c|}
		\hline
		S.No.& Terms&Nature with the variation of  $r$\\
		\hline
		1 & $\rho$ & $>0$, for $r\in(0,1)$\\
		&        & $<0$, for $r\in(1,\infty)$\\
		&        & $=0$, for $r=1$\\
		\cline{1-3}
		2 & $\rho+p_r$ &  indeterminate, for $r\in(0,44.61)$ \\
		&            & $<0$, for $r\in [44.61,58.75)\cup[63.72,\infty)$\\
		&            & $>0$, for $r\in (58.75, 63.72)$\\
		\cline{1-3}
		3 & $\rho+p_t$ & $<0$, for $r\in \{1\}\cup(2,59.25]$\\
		&            & $>0$, $r\in(0,2]\cup(59.25,\infty)-\{1\}$\\
		\cline{1-3}
		4 &  $\rho-|p_r|$ & indeterminate, for $r\in(0, 44.71)$ \\
		&               & $<0$, for $r\in [44.71, \infty)$\\
		\cline{1-3}
		5 &  $\rho-|p_t|$ & $>0$, for $r\in(0, 0.67]$ \\
		&               & $<0$, for $r\in (0.67,\infty)$\\
		\cline{1-3}
		6 &  $\triangle$ & indeterminate, for $r\in (0, 44.71)$\\
		&              & $>0$, for $r\in[44.71,58.84)\cup(63.81,\infty)$ \\
		&              & $<0$, for $r\in [58.84,63.81]$\\
		\cline{1-3}
		7 &  $\omega$& indeterminate, for $r\in (0, 44.71)$\\
		&          & $\in(0,1)$, for $r\in[44.71,58.83)$\\
		&          & $<-1$, for $r\in [58.83,\infty)$\\
		\hline
		
	\end{tabular}
\end{table}

\begin{table}[!h]
	\centering
	\caption{Summary of results in $f(R,T)$ gravity}
	\begin{tabular}{|c|c|l|}
		\hline
		S.No.& Terms&Nature with the variation of  $r$\\
		\hline
		1 & $\rho$ & $\geq0$, for $r\in(0,1]$ \& $\lambda\in(-4\pi,\infty)$\\
		&         &      $\geq0$     for $r\in[1,\infty)$ \& $\lambda\in(-\infty,-4\pi)$\\
		&        & $<0$, otherwise\\
		\cline{1-3}
		2 & $\rho+p_r$ & $\geq0$, for $r\in(0,2]$ \& $\lambda\in(-4\pi,\infty)$\\
		&         &   $\geq0$, for $r\in[2,\infty)$ \& $\lambda\in(-\infty,-4\pi)$\\
				&            & $<0$, otherwise\\
		\cline{1-3}
		3 & $\rho+p_t$ & $\geq0$, for $r\in(0,2]$ \& $\lambda\in(-4\pi,\infty)$\\
		&         &   $\geq0$, for $r\in[2,\infty)$ \& $\lambda\in(-\infty,-4\pi)$\\
		&            & $<0$, otherwise\\
		\cline{1-3}
		4 &  $\rho-|p_r|$ & $\leq0$, for $r\in(0,2]$ \& $\lambda\in(-4\pi,\infty)$\\
		&         &   $\geq0$, for $r\in[2,\infty)$ \& $\lambda\in(-\infty,-4\pi)$\\

		\cline{1-3}
		5 &  $\rho-|p_t|$ & $\geq0$, for $r\in(0,2]$ \& $\lambda\in(-4\pi,\infty)$\\
		&         &   $\geq0$, for $r\in[2,\infty)$ \& $\lambda\in(-\infty,-4\pi)$\\
		\cline{1-3}
		6 &  $\triangle$ & $<0$, for $r\in(0,2)$ \& $\lambda\in(-4\pi,\infty)$\\
		&         &   $<0$, for $r\in(2,\infty)$ \& $\lambda\in(-\infty,-4\pi)$\\
	    &         &   $=0$, for $r=2$ \& for $\lambda\neq-4\pi$\\
		\cline{1-3}
		7 &  $\omega$&  $>0$, for $r\in(0,1)$\\
		  &           & =$\infty$, for $r=1$\\
		&          & $<-1$, for $r\in (1,2)$\\
			&          & $=-1$, for $r=2$\\
				&          & $\in(-1,0)$, for $r>2$\\
		\hline
		
	\end{tabular}
\end{table}

\begin{table}[!h]
	\centering
	\caption{Summary of results in GR}
	\begin{tabular}{|c|c|l|}
		\hline
		S.No.& Terms&Nature with the variation of  $r$\\
		\hline
		1 & $\rho$ & $>0$, for $r\in(0,1)$\\
		&        & $<0$, for $r\in(1,\infty)$\\
		&        & $=0$, for $r=1$\\
		\cline{1-3}
		2 & $\rho+p_r$ & $\geq0$, for $r\in(0,2]$ \\
		&            &   $<0$, for $r\in (2,\infty)$\\
		\cline{1-3}
		3 & $\rho+p_t$ &  $\geq0$, for $r\in(0,2]$ \\
		&          &   $<0$, for $r\in (2,\infty)$\\
		\cline{1-3}
		4 &  $\rho-|p_r|$ & $<0$, for $r>0$\\
		\cline{1-3}
		5 &  $\rho-|p_t|$ & $>0$, for $r\in(0, 0.66]$ \\
		&               & $<0$, for $r\in (0.66,\infty)$\\
		\cline{1-3}
		6 &  $\triangle$ &  $<0$, for $r\in(0,2)$ \\
		&              & $>0$, for $r\in(2, \infty)$\\
		 &         &   $=0$, for $r=2$ \\
		\cline{1-3}
		7 &  $\omega$&  $>0$, for $r\in(0,1)$\\
		&           & =$\infty$, for $r=1$\\
		&          & $<-1$, for $r\in (1,2)$\\
		&          & $=-1$, for $r=2$\\
		&          & $\in(-1,0)$, for $r>2$\\
				\hline
		
	\end{tabular}
\end{table}

\noindent
\textit{Energy density}: It is observed that the energy density $\rho\geq 0$ for $r\in (0,1]$ in case of both $f(R)$ and GR. In $f(R,T)$ case, $\rho\geq 0$ for (i) $r\in (0,1]$ with $\lambda>-4\pi$ and (ii) $r\in [1,\infty)$ with $\lambda<-4\pi$. Otherwise, $\rho<0$. Thus, in both $f(R)$ gravity and GR cases, $\rho$ is positive near the throat of wormhole and negative outside the throat of wormhole. In $f(R,T)$ gravity, if $\lambda>-4\pi$, then $\rho$ is positive near the throat and negative outside the throat of wormhole and if $\lambda<-4\pi$, then $\rho$ is negative near the throat and positive outside the throat of wormhole.

\textit{NEC}: In $f(R)$ gravity and GR, NEC is  satisfied for $r\in(58.75,63.72)$ and $r\in(0,2]$ respectively. In $f(R,T)$ gravity, NEC is valid for (i) $r\in (0,2]$ with $\lambda>-4\pi$ or (ii) $r\in [2,\infty)$ with $\lambda<-4\pi$.  Thus, NEC is agreed for a large range of $r$ in $f(R,T)$ gravity, if  $\lambda<-4\pi$ is chosen. However, it is satisfied for small range of $r$ in the context of other two theories.

\textit{WEC}: In $f(R)$ case, WEC is violated everywhere. In GR case, it is satisfied for $r\in(0,1]$. Finally, in $f(R,T)$ case, it is satisfied for  (i) $r\in (0,1]$ with $\lambda>-4\pi$ or (ii) $r\in [2,\infty)$ with $\lambda<-4\pi$.  Thus,  WEC holds good in $f(R,T)$ case for $\lambda<-4\pi$.

\textit{DEC}:  In both $f(R)$ gravity and GR cases, DEC is violated throughout. However, in $f(R,T)$ gravity case, it is satisfied for $r\in [2,\infty)$, if $\lambda<-4\pi$. Thus,  DEC also holds good in $f(R,T)$ gravity for $\lambda<-4\pi$.

\textit{Anisotropy parameter $(\triangle)$}: Analyzing the results for $\triangle$, in $f(R)$ case, the geometry is repulsive for $r\in [44.71,58.84)\cup(63.81,\infty)$, attractive for $r\in [58.84,63.81]$ and not known for $r\in(0,44.71)$.  In GR case, the attractive and repulsive natures of geometry are  present for $r\in (0,2)$ and $r\in(2,\infty)$ respectively. Finally, in $f(R,T)$ gravity,  the geometry is repulsive for $r\in (0,2)$, if $\lambda>-4\pi$  or for $r\in (2,\infty)$, if $\lambda<-4\pi$, otherwise it is attractive except for $r=2$. The wormhole geometry should be  repulsive near or at the throat and attractive otherwise. This fact is fulfilled in $f(R,T)$ gravity.

\textit{Equation of state parameter $(\omega)$}: Observing the results for $\omega$, first in $f(R)$ case, there is no idea about the fluid filled for $r\in(0,44.71)$, there is ordinary fluid for $r\in[44.71,58.83)$ and phantom fluid for $r\in[58.83,\infty)$.  In $f(R,T)$ and GR cases, we have same results. In these cases, the type of fluid present is ordinary for $r\in(0,1)$, not known for $r=1$, phantom fluid for $r\in(1,2)$ and non-phantom or quintessence for $r\in[2,\infty)$.

Thus, comparing the results obtained in $f(R)$ gravity, $f(R,T)$ gravity and GR cases, it is found that the energy conditions which include NEC, WEC and DEC are satisfied only in case of $f(R,T)$ gravity for $r\in[2,\infty)$ with $\lambda<-4\pi$. The exotic matter near the throat is present in small quantity in $f(R,T)$ gravity case only. The repulsive geometric configuration which is required to keep the wormhole throat  open is found in $f(R,T)$ gravity.
Thus, all the results hold good in case of $f(R,T)$ gravity. This indicates the compatibility of $f(R, T)$ gravity in wormhole modeling.

\section{Conclusions}
In the present work,  the frameworks of $f(R)$ and $f(R, T)$ gravity theories are used for the exploration of wormhole structures. First, a novel shape function is defined in terms of the exponential function.  Then this shape function is used in field equations of both  theories and hence the energy density, null and dominant energy condition terms, equation of state parameter and anisotropy parameter are analyzed.

 In case of $f(R)$ gravity, it is found that NEC validates for a small range of radial coordinate and violates for a wide range of $r$. WEC and DEC violate everywhere. In this case, the presence of exotic matter is found in large amount, which is not an interesting result. The interesting result could be absence of exotic matter completely or presence of very less amount of exotic matter required for the construction of wormhole model.
In case of GR, it is observed that WEC is satisfied near the throat of  wormhole having attractive geometry  and dissatisfied outside the throat having repulsive geometry, which is also not preferable.

However, in case of $f(R,T)$ gravity, energy conditions, namely NEC, WEC and DEC, are satisfied for a large range of $r$ with $\lambda<-4\pi$. In this case, the wormhole geometric configuration is repulsive  near the throat. Further, very small amount of exotic matter is present near the throat. Thus, all the desirable results are found in $f(R,T)$ theory of gravity.
Eventually, we conclude that $\lambda$ plays an
important role for the existence of wormhole solutions in $f(R, T)$ gravity.
Thus, for the model undertaken, we conclude that the $f(R,T)$ theory of gravity with newly defined exponential shape function $b(r)=\frac{r}{\exp[(r-r_0)/L]}$ is a most suitable choice to describe the existence of wormhole solutions filled with very less amount of exotic matter near the throat of  wormhole.

\section{Appendix A}

In $f(R)$ gravity, the wormhole solutions are
\begin{eqnarray}
	\rho&=&-\frac{1}{16 R_c^2 L^4 r^6}\Bigg[e^{-\frac{3 r}{L}} \text{sech}^4\left(\frac{2 e^{\frac{r_0-r}{L}} (L-r)}{R_c L r^2}\right) \left(2 R_c^3 e^{\frac{3 r}{L}} L^4 \mu  \sinh \left(\frac{4 e^{\frac{r_0-r}{L}} (L-r)}{R_c L r^2}\right) r^6+R_c^3 e^{\frac{3 r}{L}}  \right.\nonumber\\
	&\times&\left.L^4 \mu \sinh \left(\frac{8 e^{\frac{r_0-r}{L}} (L-r)}{R_c L r^2}\right) r^6+6 R_c^2 e^{\frac{2 r+r_0}{L}} L^3 r^5-32 R_c e^{\frac{2 r+r_0}{L}} L \mu  \sinh \left(\frac{4 e^{\frac{r_0-r}{L}} (L-r)}{R_c L r^2}\right) r^5\right.\nonumber\\
	&+&\left.16 R_c e^{\frac{r+2 r_0}{L}} L \mu  \sinh \left(\frac{4 e^{\frac{r_0-r}{L}} (L-r)}{R_c L r^2}\right) r^5-6 R_c^2 e^{\frac{2 r+r_0}{L}} L^4 r^4-256 e^{\frac{3 r_0}{L}} \mu  r^4+256 e^{\frac{r+2 r_0}{L}} \mu  r^4\right.\nonumber\\
	&-&\left.2 R_c^2 e^{\frac{2 r+r_0}{L}} L^3 (L-r) \cosh \left(\frac{8 e^{\frac{r_0-r}{L}} (L-r)}{R_c L r^2}\right) r^4+32 R_c e^{\frac{2 r+r_0}{L}} L^2 \mu  \sinh \left(\frac{4 e^{\frac{r_0-r}{L}} (L-r)}{R_c L r^2}\right) r^4 \right.\nonumber\\
	&-&\left. 32 R_c e^{\frac{r+2 r_0}{L}} L^2 \mu \sinh \left(\frac{4 e^{\frac{r_0-r}{L}} (L-r)}{R_c L r^2}\right) r^4+64 R_c e^{\frac{2 r+r_0}{L}} L^3 \mu  \sinh \left(\frac{4 e^{\frac{r_0-r}{L}} (L-r)}{R_c L r^2}\right) r^3 \right.\nonumber\\
	&-&\left. 32 R_c e^{\frac{r+2 r_0}{L}} L^3 \mu\sinh \left(\frac{4 e^{\frac{r_0-r}{L}} (L-r)}{R_c L r^2}\right) r^3+1024 e^{\frac{3 r_0}{L}} L^2 \mu  r^2-1024 e^{\frac{r+2 r_0}{L}} L^2 \mu  r^2 \right.\nonumber\\
	&+&\left.64 R_c e^{\frac{2 r+r_0}{L}} L^4 \mu \sinh \left(\frac{4 e^{\frac{r_0-r}{L}} (L-r)}{R_c L r^2}\right) r^2-64 R_c e^{\frac{r+2 r_0}{L}} L^4 \mu  \sinh \left(\frac{4 e^{\frac{r_0-r}{L}} (L-r)}{R_c L r^2}\right) r^2\right.\nonumber\\
	&-&\left.-1024 e^{\frac{3 r_0}{L}} L^4 \mu +1024 e^{\frac{r+2 r_0}{L}} L^4 \mu +8 e^{r_0/L} \left(16 e^{r_0/L} \left(-e^{r/L}+e^{r_0/L}\right) \left(r^2-2 L^2\right)^2 \mu \right.\right.\nonumber\\
	&-& \left.\left.-R_c^2 e^{\frac{2 r}{L}} L^3 (L-r) r^4\right)\cosh \left(\frac{4 e^{\frac{r_0-r}{L}} (L-r)}{R_c L r^2}\right)\right)\Bigg]\label{rr1}
\end{eqnarray}

\begin{eqnarray}
	p_r&=&\frac{1}{R_c L^2 r^4}\Bigg[\mu  e^{\frac{r_0-2 r}{L}} \left(R_c L r^3 e^{r/L}-8 \left(2 L^2-r^2\right) \left(e^{r/L}-e^{r_0/L}\right) \tanh \left(\frac{2 (L-r) e^{\frac{r_0-r}{L}}}{R_c L r^2}\right)\right)\nonumber\\
	&\times& \text{sech}^2\left(\frac{2 (L-r) e^{\frac{r_0-r}{L}}}{R_c L r^2}\right)\Bigg]+\frac{1}{2} R_c \mu  \tanh \left(\frac{2 (L-r) e^{\frac{r_0-r}{L}}}{R_c L r^2}\right)-\frac{e^{\frac{r_0-r}{L}}}{r^2}\label{prr1}
\end{eqnarray}
\begin{eqnarray}
	p_t&=&\frac{1}{16 R_c^2 L^4 r^6}\Bigg[e^{-\frac{3 r}{L}} \text{sech}^4\left(\frac{2 (L-r) e^{\frac{r_0-r}{L}}}{R_c L r^2}\right) \left(2 R_c^3 \mu  L^4 r^6 e^{\frac{3 r}{L}} \sinh \left(\frac{4 (L-r) e^{\frac{r_0-r}{L}}}{R_c L r^2}\right)\right.\nonumber\\
	&+&\left. R_c^3 \mu  L^4 r^6 e^{\frac{3 r}{L}} \sinh \left(\frac{8 (L-r) e^{\frac{r_0-r}{L}}}{R_c L r^2}\right)-8 R_c^2 \mu  L^4 r^4 e^{\frac{2 r+r_0}{L}}+4 R_c^2 \mu  L^3 r^5 e^{\frac{2 r+r_0}{L}}\right.\nonumber\\
	&+&\left.3 R_c^2 L^3 r^5 e^{\frac{2 r+r_0}{L}}+R_c^2 L^3 r^5 e^{\frac{2 r+r_0}{L}} \cosh \left(\frac{8 (L-r) e^{\frac{r_0-r}{L}}}{R_c L r^2}\right)+4 e^{r_0/L} \left(32 \mu  \left(r^2-2 L^2\right)^2\right.\right.\nonumber
\end{eqnarray}
\begin{eqnarray}
	&\times&\left.\left.  e^{r_0/L} \left(e^{r_0/L}-e^{r/L}\right)-R_c^2 L^3 r^4 e^{\frac{2 r}{L}} (2 \mu  L-(\mu +1) r)\right) \cosh \left(\frac{4 (L-r) e^{\frac{r_0-r}{L}}}{R_c L r^2}\right)+128 R_c \mu  L^4 r^2 \right.\nonumber\\
	&\times&\left. 
	e^{\frac{2 r+r_0}{L}} \sinh \left(\frac{4 (L-r) e^{\frac{r_0-r}{L}}}{R_c L r^2}\right)-128 R_c \mu  L^4 r^2
	e^{\frac{r+2 r_0}{L}} \sinh \left(\frac{4 (L-r) e^{\frac{r_0-r}{L}}}{R_c L r^2}\right)+64 R_c \mu  L^3 r^3 e^{\frac{2 r+r_0}{L}}\right.\nonumber\\
	&\times&\left. 
	\sinh \left(\frac{4 (L-r) e^{\frac{r_0-r}{L}}}{R_c L r^2}\right)-32 R_c \mu  L^3 r^3 e^{\frac{r+2 r_0}{L}} \sinh \left(\frac{4 (L-r) e^{\frac{r_0-r}{L}}}{R_c L r^2}\right)-32 R_c \mu  L r^5 e^{\frac{2 r+r_0}{L}}\right.\nonumber\\
	&\times&\left.  \sinh \left(\frac{4 (L-r) e^{\frac{r_0-r}{L}}}{R_c L r^2}\right)+16 R_c \mu  L r^5 e^{\frac{r+2 r_0}{L}} \sinh \left(\frac{4 (L-r) e^{\frac{r_0-r}{L}}}{R_c L r^2}\right)+1024 \mu  L^4 e^{\frac{r+2 r_0}{L}}\right.\nonumber\\
	&-&\left.  1024 \mu  L^4  e^{\frac{3 r_0}{L}}+1024 \mu  L^2 r^2 e^{\frac{3 r_0}{L}}-1024 \mu  L^2 r^2 e^{\frac{r+2 r_0}{L}}-256 \mu  r^4 e^{\frac{3 r_0}{L}}+256 \mu  r^4 e^{\frac{r+2 r_0}{L}}\right)\Bigg]
\end{eqnarray}

\noindent
From Equations (42), (43) and (44),
\begin{eqnarray}
	\rho +  p_r&=&-\frac{1}{16 R_c^2 L^4 r^6}\Bigg[e^{-\frac{3 r}{L}} \text{sech}^4\left(\frac{2 e^{\frac{r_0-r}{L}} (L-r)}{R_c L r^2}\right) \left(2 R_c^3 e^{\frac{3 r}{L}} L^4 \mu  \sinh \left(\frac{4 e^{\frac{r_0-r}{L}} (L-r)}{R_c L r^2}\right) r^6 \right.\nonumber\\
	&+&\left.R_c^3 e^{\frac{3 r}{L}} L^4 \mu \sinh \left(\frac{8 e^{\frac{r_0-r}{L}} (L-r)}{R_c L r^2}\right) r^6+6 R_c^2 e^{\frac{2 r+r_0}{L}} L^3 r^5-32 R_c e^{\frac{2 r+r_0}{L}} L \mu \right.
	\nonumber\\
	&\times& \left.\sinh \left(\frac{4 e^{\frac{r_0-r}{L}} (L-r)}{R_c L r^2}\right) r^5+16 R_c e^{\frac{r+2 r_0}{L}} L \mu  \sinh \left(\frac{4 e^{\frac{r_0-r}{L}} (L-r)}{R_c L r^2}\right) r^5-6 R_c^2 e^{\frac{2 r+r_0}{L}} L^4 r^4\right.\nonumber\\
	&\times&\left.
	-256 e^{\frac{3 r_0}{L}} \mu  r^4+256 e^{\frac{r+2 r_0}{L}} \mu  r^4-2 R_c^2 e^{\frac{2 r+r_0}{L}} L^3 (L-r) \cosh \left(\frac{8 e^{\frac{r_0-r}{L}} (L-r)}{R_c L r^2}\right) r^4\right.\nonumber\\
	&+&\left.32 R_c e^{\frac{2 r+r_0}{L}} L^2 \mu  \sinh \left(\frac{4 e^{\frac{r_0-r}{L}} (L-r)}{R_c L r^2}\right) r^4-32 R_c e^{\frac{r+2 r_0}{L}} L^2 \mu\sinh \left(\frac{4 e^{\frac{r_0-r}{L}} (L-r)}{R_c L r^2}\right) r^4 \right.\nonumber\\
	&+&\left. 64 R_c e^{\frac{2 r+r_0}{L}} L^3 \mu  \sinh \left(\frac{4 e^{\frac{r_0-r}{L}} (L-r)}{R_c L r^2}\right) r^3-32 R_c e^{\frac{r+2 r_0}{L}} L^3 \mu  \sinh \left(\frac{4 e^{\frac{r_0-r}{L}} (L-r)}{R_c L r^2}\right) r^3\right.\nonumber\\
	&+&\left.1024 e^{\frac{3 r_0}{L}} L^2 \mu  r^2-1024 e^{\frac{r+2 r_0}{L}} L^2 \mu  r^2+64 R_c e^{\frac{2 r+r_0}{L}} L^4 \mu\sinh \left(\frac{4 e^{\frac{r_0-r}{L}} (L-r)}{R_c L r^2}\right) r^2 \right.\nonumber\\
	&-&\left. 64 R_c e^{\frac{r+2 r_0}{L}} L^4 \mu  \sinh \left(\frac{4 e^{\frac{r_0-r}{L}} (L-r)}{R_c L r^2}\right) r^2-1024 e^{\frac{3 r_0}{L}} L^4 \mu +1024e^{\frac{r+2 r_0}{L}} L^4 \mu\right.\nonumber\\
	&+&\left. 8 e^{r_0/L} \left(16 e^{r_0/L} \left(-e^{r/L}+e^{r_0/L}\right) \left(r^2-2 L^2\right)^2 \mu -R_c^2 e^{\frac{2 r}{L}} L^3 (L-r) r^4\right)\right.\nonumber\\
	&\times& \left.\cosh \left(\frac{4 e^{\frac{r_0-r}{L}} (L-r)}{R_c L r^2}\right)\right)\Bigg]+\frac{1}{R_c L^2 r^4}\Bigg[\mu  e^{\frac{r_0-2 r}{L}} \left(R_c L r^3 e^{r/L}-8 \left(2 L^2-r^2\right) \right.\nonumber\\
	&\times&  \left.\left(e^{r/L}-e^{r_0/L}\right)\tanh \left(\frac{2 (L-r) e^{\frac{r_0-r}{L}}}{R_c L r^2}\right)\right) \text{sech}^2\left(\frac{2 (L-r) e^{\frac{r_0-r}{L}}}{R_c L r^2}\right)\Bigg]\nonumber
\end{eqnarray}

\begin{eqnarray}
	&+& 
	\frac{1}{2} R_c \mu  \tanh \left(\frac{2 (L-r) e^{\frac{r_0-r}{L}}}{R_c L r^2}\right)-\frac{e^{\frac{r_0-r}{L}}}{r^2}
\end{eqnarray}

\begin{eqnarray}
	\rho +  p_t&=&-\frac{1}{4 R_c L^2 r^4}\Bigg[e^{\frac{\text{r0}-2 r}{L}} \text{sech}^2\left(\frac{2 (L-r) e^{\frac{\text{r0}-r}{L}}}{R_c L r^2}\right) \left(-16 \mu  \left(2 L^2-r^2\right) \left(e^{r/L}-e^{\text{r0}/L}\right)\right.\nonumber\\
	&\times& \left. \tanh \left(\frac{2 (L-r) e^{\frac{\text{r0}-r}{L}}}{R_c L r^2}\right)+R_c (2 \mu -1) L r^2 e^{r/L} (2 L-r)-R_c L r^2 e^{r/L} (2 L-r)\right.\nonumber\\
	&\times& \left.
	\cosh \left(\frac{4 (L-r) e^{\frac{\text{r0}-r}{L}}}{R_c L r^2}\right)\right)\Bigg]
\end{eqnarray}

%

\begin{eqnarray}
	\rho -|p_r|&=&-\frac{1}{16 R_c^2 L^4 r^6}\Bigg[e^{-\frac{3 r}{L}} \text{sech}^4\left(\frac{2 e^{\frac{r_0-r}{L}} (L-r)}{R_c L r^2}\right) \left(2 R_c^3 e^{\frac{3 r}{L}} L^4 \mu  \sinh \left(\frac{4 e^{\frac{r_0-r}{L}} (L-r)}{R_c L r^2}\right) r^6 \right.\nonumber\\
	&+&\left.R_c^3 e^{\frac{3 r}{L}} L^4 \mu \sinh \left(\frac{8 e^{\frac{r_0-r}{L}} (L-r)}{R_c L r^2}\right) r^6+6 R_c^2 e^{\frac{2 r+r_0}{L}} L^3 r^5-32 R_c e^{\frac{2 r+r_0}{L}} L \mu \right.\nonumber\\
	&\times&\left. \sinh \left(\frac{4 e^{\frac{r_0-r}{L}} (L-r)}{R_c L r^2}\right) r^5+16 R_c e^{\frac{r+2 r_0}{L}} L \mu  \sinh \left(\frac{4 e^{\frac{r_0-r}{L}} (L-r)}{R_c L r^2}\right) r^5-6 R_c^2 e^{\frac{2 r+r_0}{L}} L^4 r^4\right.\nonumber\\
	&-&\left.256 e^{\frac{3 r_0}{L}} \mu  r^4+256 e^{\frac{r+2 r_0}{L}} \mu  r^4-2 R_c^2 e^{\frac{2 r+r_0}{L}} L^3 (L-r) \cosh \left(\frac{8 e^{\frac{r_0-r}{L}} (L-r)}{R_c L r^2}\right) r^4\right.\nonumber\\
	&+&\left.32 R_c e^{\frac{2 r+r_0}{L}} L^2 \mu  \sinh \left(\frac{4 e^{\frac{r_0-r}{L}} (L-r)}{R_c L r^2}\right) r^4-32 R_c e^{\frac{r+2 r_0}{L}} L^2 \mu \sinh \left(\frac{4 e^{\frac{r_0-r}{L}} (L-r)}{R_c L r^2}\right) r^4\right.\nonumber\\
	&+&\left.64 R_c e^{\frac{2 r+r_0}{L}} L^3 \mu  \sinh \left(\frac{4 e^{\frac{r_0-r}{L}} (L-r)}{R_c L r^2}\right) r^3-32 R_c e^{\frac{r+2 r_0}{L}} L^3 \mu \sinh \left(\frac{4 e^{\frac{r_0-r}{L}} (L-r)}{R_c L r^2}\right) r^3
	\right.\nonumber\\
	&+&\left. 1024 e^{\frac{3 r_0}{L}} L^2 \mu  r^2-1024 e^{\frac{r+2 r_0}{L}} L^2 \mu  r^2+64 R_c e^{\frac{2 r+r_0}{L}} L^4 \mu\sinh \left(\frac{4 e^{\frac{r_0-r}{L}} (L-r)}{R_c L r^2}\right) r^2\right.\nonumber\\
	&-&\left.64 R_c e^{\frac{r+2 r_0}{L}} L^4 \mu  \sinh \left(\frac{4 e^{\frac{r_0-r}{L}} (L-r)}{R_c L r^2}\right) r^2-1024 e^{\frac{3 r_0}{L}} L^4 \mu +1024 e^{\frac{r+2 r_0}{L}} L^4 \mu +8 e^{r_0/L} \right.\nonumber\\
	&\times& \left.\left(16 e^{r_0/L} \left(-e^{r/L}+e^{r_0/L}\right) \left(r^2-2 L^2\right)^2 \mu -R_c^2 e^{\frac{2 r}{L}} L^3 (L-r) r^4\right)\cosh \left(\frac{4 e^{\frac{r_0-r}{L}} (L-r)}{R_c L r^2}\right)\right)\Bigg]\nonumber\\
	&-&\Bigg|\frac{1}{R_c L^2 r^4}\Bigg[\mu  e^{\frac{r_0-2 r}{L}} \left(R_c L r^3 e^{r/L}-8 \left(2 L^2-r^2\right) \left(e^{r/L}-e^{r_0/L}\right) \tanh \left(\frac{2 (L-r) e^{\frac{r_0-r}{L}}}{R_c L r^2}\right)\right)\nonumber\\
	&\times& \text{sech}^2\left(\frac{2 (L-r) e^{\frac{r_0-r}{L}}}{R_c L r^2}\right)\Bigg]+\frac{1}{2} R_c \mu  \tanh \left(\frac{2 (L-r) e^{\frac{r_0-r}{L}}}{R_c L r^2}\right)-\frac{e^{\frac{r_0-r}{L}}}{r^2}\Bigg|
\end{eqnarray}

\begin{eqnarray}
	\rho -|p_t|&=&-\frac{1}{16 R_c^2 L^4 r^6}\Bigg[e^{-\frac{3 r}{L}} \text{sech}^4\left(\frac{2 e^{\frac{r_0-r}{L}} (L-r)}{R_c L r^2}\right) \left(2 R_c^3 e^{\frac{3 r}{L}} L^4 \mu  \sinh \left(\frac{4 e^{\frac{r_0-r}{L}} (L-r)}{R_c L r^2}\right) r^6 \right.\nonumber\\
	&+&\left.R_c^3 e^{\frac{3 r}{L}} L^4 \mu \sinh \left(\frac{8 e^{\frac{r_0-r}{L}} (L-r)}{R_c L r^2}\right) r^6+6 R_c^2 e^{\frac{2 r+r_0}{L}} L^3 r^5-32 R_c e^{\frac{2 r+r_0}{L}} L \mu \right.\nonumber\\
	&\times&\left. \sinh \left(\frac{4 e^{\frac{r_0-r}{L}} (L-r)}{R_c L r^2}\right) r^5+16 R_c e^{\frac{r+2 r_0}{L}} L \mu  \sinh \left(\frac{4 e^{\frac{r_0-r}{L}} (L-r)}{R_c L r^2}\right) r^5-6 R_c^2 e^{\frac{2 r+r_0}{L}} L^4 r^4\right.\nonumber\\
	&-&\left.256 e^{\frac{3 r_0}{L}} \mu  r^4+256 e^{\frac{r+2 r_0}{L}} \mu  r^4-2 R_c^2 e^{\frac{2 r+r_0}{L}} L^3 (L-r) \cosh \left(\frac{8 e^{\frac{r_0-r}{L}} (L-r)}{R_c L r^2}\right) r^4\right.\nonumber\\
	&+&\left.32 R_c e^{\frac{2 r+r_0}{L}} L^2 \mu  \sinh \left(\frac{4 e^{\frac{r_0-r}{L}} (L-r)}{R_c L r^2}\right) r^4-32 R_c e^{\frac{r+2 r_0}{L}} L^2 \mu \sinh \left(\frac{4 e^{\frac{r_0-r}{L}} (L-r)}{R_c L r^2}\right) r^4\right.\nonumber\\
	&+&\left.64 R_c e^{\frac{2 r+r_0}{L}} L^3 \mu  \sinh \left(\frac{4 e^{\frac{r_0-r}{L}} (L-r)}{R_c L r^2}\right) r^3-32 R_c e^{\frac{r+2 r_0}{L}} L^3 \mu \sinh \left(\frac{4 e^{\frac{r_0-r}{L}} (L-r)}{R_c L r^2}\right) r^3
	\right.\nonumber\\
	&+&\left. 1024 e^{\frac{3 r_0}{L}} L^2 \mu  r^2-1024 e^{\frac{r+2 r_0}{L}} L^2 \mu  r^2+64 R_c e^{\frac{2 r+r_0}{L}} L^4 \mu\sinh \left(\frac{4 e^{\frac{r_0-r}{L}} (L-r)}{R_c L r^2}\right) r^2\right.\nonumber\\
	&-&\left.64 R_c e^{\frac{r+2 r_0}{L}} L^4 \mu  \sinh \left(\frac{4 e^{\frac{r_0-r}{L}} (L-r)}{R_c L r^2}\right) r^2-1024 e^{\frac{3 r_0}{L}} L^4 \mu +1024 e^{\frac{r+2 r_0}{L}} L^4 \mu +8 e^{r_0/L} \right.\nonumber\\
	&\times& \left.\left(16 e^{r_0/L} \left(-e^{r/L}+e^{r_0/L}\right) \left(r^2-2 L^2\right)^2 \mu -R_c^2 e^{\frac{2 r}{L}} L^3 (L-r) r^4\right)\cosh \left(\frac{4 e^{\frac{r_0-r}{L}} (L-r)}{R_c L r^2}\right)\right)\Bigg]\nonumber\\
	&-&\Bigg|\frac{1}{16 R_c^2 L^4 r^6}\Bigg[e^{-\frac{3 r}{L}} \text{sech}^4\left(\frac{2 (L-r) e^{\frac{r_0-r}{L}}}{R_c L r^2}\right) \left(2 R_c^3 \mu  L^4 r^6 e^{\frac{3 r}{L}} \sinh \left(\frac{4 (L-r) e^{\frac{r_0-r}{L}}}{R_c L r^2}\right)\right.\nonumber\\
	&+&\left. R_c^3 \mu  L^4 r^6 e^{\frac{3 r}{L}} \sinh \left(\frac{8 (L-r) e^{\frac{r_0-r}{L}}}{R_c L r^2}\right)-8 R_c^2 \mu  L^4 r^4 e^{\frac{2 r+r_0}{L}}+4 R_c^2 \mu  L^3 r^5 e^{\frac{2 r+r_0}{L}}\right.\nonumber\\
	&+&\left.3 R_c^2 L^3 r^5 e^{\frac{2 r+r_0}{L}}+R_c^2 L^3 r^5 e^{\frac{2 r+r_0}{L}} \cosh \left(\frac{8 (L-r) e^{\frac{r_0-r}{L}}}{R_c L r^2}\right)+4 e^{r_0/L} \left(32 \mu  \left(r^2-2 L^2\right)^2\right.\right.\nonumber\\
	&\times&\left.\left.  e^{r_0/L} \left(e^{r_0/L}-e^{r/L}\right)-R_c^2 L^3 r^4 e^{\frac{2 r}{L}} (2 \mu  L-(\mu +1) r)\right) \cosh \left(\frac{4 (L-r) e^{\frac{r_0-r}{L}}}{R_c L r^2}\right) \right.\nonumber\\
	&+&\left. 
	128 R_c \mu  L^4 r^2e^{\frac{2 r+r_0}{L}} \sinh \left(\frac{4 (L-r) e^{\frac{r_0-r}{L}}}{R_c L r^2}\right)-128 R_c \mu  L^4 r^2
	e^{\frac{r+2 r_0}{L}} \sinh \left(\frac{4 (L-r) e^{\frac{r_0-r}{L}}}{R_c L r^2}\right)\right.\nonumber\\
	&+&\left. +64 R_c \mu  L^3 r^3 e^{\frac{2 r+r_0}{L}} 
	\sinh \left(\frac{4 (L-r) e^{\frac{r_0-r}{L}}}{R_c L r^2}\right)-32 R_c \mu  L^3 r^3 e^{\frac{r+2 r_0}{L}} \sinh \left(\frac{4 (L-r) e^{\frac{r_0-r}{L}}}{R_c L r^2}\right)\right.\nonumber\\
	&-&\left.32 R_c \mu  L r^5 e^{\frac{2 r+r_0}{L}}  \sinh \left(\frac{4 (L-r) e^{\frac{r_0-r}{L}}}{R_c L r^2}\right)+16 R_c \mu  L r^5 e^{\frac{r+2 r_0}{L}} \sinh \left(\frac{4 (L-r) e^{\frac{r_0-r}{L}}}{R_c L r^2}\right)\right.\nonumber
\end{eqnarray}

\begin{eqnarray}
	&+&\left.1024 \mu  L^4 e^{\frac{r+2 r_0}{L}}-  1024 \mu  L^4  e^{\frac{3 r_0}{L}}+1024 \mu  L^2 r^2 e^{\frac{3 r_0}{L}}-1024 \mu  L^2 r^2 e^{\frac{r+2 r_0}{L}}-256 \mu  r^4 e^{\frac{3 r_0}{L}}\right.\nonumber\\
	&+&\left. 
	+256 \mu  r^4 e^{\frac{r+2 r_0}{L}}\right)\Bigg]
	\Bigg|
\end{eqnarray}

\begin{eqnarray}
	p_t-p_r&=&\frac{1}{16 R_c^2 L^4 r^6}\Bigg[e^{-\frac{3 r}{L}} \text{sech}^4\left(\frac{2 (L-r) e^{\frac{r_0-r}{L}}}{R_c L r^2}\right) \left(2 R_c^3 \mu  L^4 r^6 e^{\frac{3 r}{L}} \sinh \left(\frac{4 (L-r) e^{\frac{r_0-r}{L}}}{R_c L r^2}\right)\right.\nonumber\\
	&+&\left. R_c^3 \mu  L^4 r^6 e^{\frac{3 r}{L}} \sinh \left(\frac{8 (L-r) e^{\frac{r_0-r}{L}}}{R_c L r^2}\right)-8 R_c^2 \mu  L^4 r^4 e^{\frac{2 r+r_0}{L}}+4 R_c^2 \mu  L^3 r^5 e^{\frac{2 r+r_0}{L}}\right.\nonumber\\
	&+&\left.3 R_c^2 L^3 r^5 e^{\frac{2 r+r_0}{L}}+R_c^2 L^3 r^5 e^{\frac{2 r+r_0}{L}} \cosh \left(\frac{8 (L-r) e^{\frac{r_0-r}{L}}}{R_c L r^2}\right)+4 e^{r_0/L} \left(32 \mu  \left(r^2-2 L^2\right)^2\right.\right.\nonumber\\
	&\times&\left.\left.  e^{r_0/L} \left(e^{r_0/L}-e^{r/L}\right)-R_c^2 L^3 r^4 e^{\frac{2 r}{L}} (2 \mu  L-(\mu +1) r)\right) \cosh \left(\frac{4 (L-r) e^{\frac{r_0-r}{L}}}{R_c L r^2}\right)+128 R_c \mu  L^4 r^2 \right.\nonumber\\
	&\times&\left. 
	e^{\frac{2 r+r_0}{L}} \sinh \left(\frac{4 (L-r) e^{\frac{r_0-r}{L}}}{R_c L r^2}\right)-128 R_c \mu  L^4 r^2
	e^{\frac{r+2 r_0}{L}} \sinh \left(\frac{4 (L-r) e^{\frac{r_0-r}{L}}}{R_c L r^2}\right)+64 R_c \mu  L^3 r^3 e^{\frac{2 r+r_0}{L}}\right.\nonumber\\
	&\times&\left. 
	\sinh \left(\frac{4 (L-r) e^{\frac{r_0-r}{L}}}{R_c L r^2}\right)-32 R_c \mu  L^3 r^3 e^{\frac{r+2 r_0}{L}} \sinh \left(\frac{4 (L-r) e^{\frac{r_0-r}{L}}}{R_c L r^2}\right)-32 R_c \mu  L r^5 e^{\frac{2 r+r_0}{L}}\right.\nonumber\\
	&\times&\left.  \sinh \left(\frac{4 (L-r) e^{\frac{r_0-r}{L}}}{R_c L r^2}\right)+16 R_c \mu  L r^5 e^{\frac{r+2 r_0}{L}} \sinh \left(\frac{4 (L-r) e^{\frac{r_0-r}{L}}}{R_c L r^2}\right)+1024 \mu  L^4 e^{\frac{r+2 r_0}{L}}\right.\nonumber\\
	&-&\left.  1024 \mu  L^4  e^{\frac{3 r_0}{L}}+1024 \mu  L^2 r^2 e^{\frac{3 r_0}{L}}-1024 \mu  L^2 r^2 e^{\frac{r+2 r_0}{L}}-256 \mu  r^4 e^{\frac{3 r_0}{L}}+256 \mu  r^4 e^{\frac{r+2 r_0}{L}}\right)\Bigg]\nonumber\\
	&-&\frac{1}{R_c L^2 r^4}\Bigg[\mu  e^{\frac{r_0-2 r}{L}} \left(R_c L r^3 e^{r/L}-8 \left(2 L^2-r^2\right) \left(e^{r/L}-e^{r_0/L}\right) \tanh \left(\frac{2 (L-r) e^{\frac{r_0-r}{L}}}{R_c L r^2}\right)\right)\nonumber\\
	&\times& \text{sech}^2\left(\frac{2 (L-r) e^{\frac{r_0-r}{L}}}{R_c L r^2}\right)\Bigg]+\frac{1}{2} R_c \mu  \tanh \left(\frac{2 (L-r) e^{\frac{r_0-r}{L}}}{R_c L r^2}\right)-\frac{e^{\frac{r_0-r}{L}}}{r^2}
\end{eqnarray}

\begin{eqnarray}
	\frac{p_r}{\rho}&=&\frac{1}{R_c L^2 r^4}\Bigg[\mu  e^{\frac{r_0-2 r}{L}} \left(R_c L r^3 e^{r/L}-8 \left(2 L^2-r^2\right) \left(e^{r/L}-e^{r_0/L}\right) \tanh \left(\frac{2 (L-r) e^{\frac{r_0-r}{L}}}{R_c L r^2}\right)\right)\nonumber\\
	&\times& \text{sech}^2\left(\frac{2 (L-r) e^{\frac{r_0-r}{L}}}{R_c L r^2}\right)\Bigg]+\frac{1}{2} R_c \mu  \tanh \left(\frac{2 (L-r) e^{\frac{r_0-r}{L}}}{R_c L r^2}\right)-\frac{e^{\frac{r_0-r}{L}}}{r^2}\nonumber\\
	&\div& -\frac{1}{16 R_c^2 L^4 r^6}\Bigg[e^{-\frac{3 r}{L}} \text{sech}^4\left(\frac{2 e^{\frac{r_0-r}{L}} (L-r)}{R_c L r^2}\right) \left(2 R_c^3 e^{\frac{3 r}{L}} L^4 \mu  \sinh \left(\frac{4 e^{\frac{r_0-r}{L}} (L-r)}{R_c L r^2}\right) r^6+R_c^3 e^{\frac{3 r}{L}} L^4 \mu \right.\nonumber\\
	&\times&\left. \sinh \left(\frac{8 e^{\frac{r_0-r}{L}} (L-r)}{R_c L r^2}\right) r^6+6 R_c^2 e^{\frac{2 r+r_0}{L}} L^3 r^5-32 R_c e^{\frac{2 r+r_0}{L}} L \mu  \sinh \left(\frac{4 e^{\frac{r_0-r}{L}} (L-r)}{R_c L r^2}\right) r^5+16 R_c e^{\frac{r+2 r_0}{L}}\right.\nonumber
	\end{eqnarray}
	
	\begin{eqnarray}
	&\times&\left. L \mu  \sinh \left(\frac{4 e^{\frac{r_0-r}{L}} (L-r)}{R_c L r^2}\right) r^5-6 R_c^2 e^{\frac{2 r+r_0}{L}} L^4 r^4-256 e^{\frac{3 r_0}{L}} \mu  r^4+256 e^{\frac{r+2 r_0}{L}} \mu  r^4-2 R_c^2 e^{\frac{2 r+r_0}{L}} L^3\right.\nonumber\\
	&\times&\left. (L-r) \cosh \left(\frac{8 e^{\frac{r_0-r}{L}} (L-r)}{R_c L r^2}\right) r^4+32 R_c e^{\frac{2 r+r_0}{L}} L^2 \mu  \sinh \left(\frac{4 e^{\frac{r_0-r}{L}} (L-r)}{R_c L r^2}\right) r^4-32 R_c e^{\frac{r+2 r_0}{L}} L^2 \mu \right.\nonumber\\
	&\times&\left. \sinh \left(\frac{4 e^{\frac{r_0-r}{L}} (L-r)}{R_c L r^2}\right) r^4+64 R_c e^{\frac{2 r+r_0}{L}} L^3 \mu  \sinh \left(\frac{4 e^{\frac{r_0-r}{L}} (L-r)}{R_c L r^2}\right) r^3-32 R_c e^{\frac{r+2 r_0}{L}} L^3 \mu \right.\nonumber\\
	&\times&\left. \sinh \left(\frac{4 e^{\frac{r_0-r}{L}} (L-r)}{R_c L r^2}\right) r^3+1024 e^{\frac{3 r_0}{L}} L^2 \mu  r^2-1024 e^{\frac{r+2 r_0}{L}} L^2 \mu  r^2+64 R_c e^{\frac{2 r+r_0}{L}} L^4 \mu \right.\nonumber\\
	&\times&\left. \sinh \left(\frac{4 e^{\frac{r_0-r}{L}} (L-r)}{R_c L r^2}\right) r^2-64 R_c e^{\frac{r+2 r_0}{L}} L^4 \mu  \sinh \left(\frac{4 e^{\frac{r_0-r}{L}} (L-r)}{R_c L r^2}\right) r^2-1024 e^{\frac{3 r_0}{L}} L^4 \mu +1024\right.\nonumber\\
	&\times&\left. e^{\frac{r+2 r_0}{L}} L^4 \mu +8 e^{r_0/L} \left(16 e^{r_0/L} \left(-e^{r/L}+e^{r_0/L}\right) \left(r^2-2 L^2\right)^2 \mu -R_c^2 e^{\frac{2 r}{L}} L^3 (L-r) r^4\right)\right.\nonumber\\
	&\times& \left.\cosh \left(\frac{4 e^{\frac{r_0-r}{L}} (L-r)}{R_c L r^2}\right)\right)\Bigg]
\end{eqnarray}\\

\noindent
\textbf{Acknowledgement:} The first author G. C. Samanta is extremely thankful to Council of Scientific and Industrial Research (CSIR),
Govt. of India for providing financial support \textbf{(Ref. No. 25(0260)/17/EMR-II)} to carry out the research work. Furthermore, the work of KB was supported in part by the JSPS KAKENHI Grant Number JP
25800136 and Competitive Research Funds for Fukushima University Faculty
(18RI009). The authors are very much grateful to the anonymous Reviewer and Editor for their meaningful comments and suggestions to improve the quality of research.

\end{document}